\documentclass[a4paper,12pt]{article}

\usepackage{lmodern}            
\usepackage[T1]{fontenc}        
\usepackage[utf8]{inputenc}     
\usepackage{textcomp}           
\usepackage{a4wide}             

\usepackage{amsmath}            
\usepackage{amssymb}            
\usepackage{amsfonts}           
\usepackage{units}              
\usepackage{slashed}            
\usepackage{fleqn}

\usepackage{caption}            
\usepackage{subfig}             
\usepackage{cite}               

\usepackage{graphicx}           
\usepackage{psfrag}             
\usepackage[dvipsnames]{xcolor} 
\usepackage{tikz}               

\usepackage{booktabs}           
\usepackage{multirow}           
\usepackage{bigstrut}           

\usepackage{hyperref}           
\usepackage{subfig}

\usepackage{amsmath,amssymb,amsfonts}	
\usepackage{color}
\usepackage{amsmath}
\usepackage{amsfonts}
\usepackage{amssymb}
\usepackage{graphicx}
\usepackage{slashed}            
\usepackage{xspace} 		 
\usepackage{slashbox,pict2e}
\usepackage{makecell}

\numberwithin{equation}{section} 
\usepackage{tikz}
\usetikzlibrary{arrows,shapes}
\usetikzlibrary{trees}
\usetikzlibrary{matrix,arrows} 				
\usetikzlibrary{positioning}				
\usetikzlibrary{calc,through}				
\usetikzlibrary{decorations.pathreplacing}  
\usepackage{pgffor}							

\usetikzlibrary{decorations.pathmorphing}	
\usetikzlibrary{decorations.markings}
\tikzset{
    vector/.style={decorate, decoration={snake}, draw},
	provector/.style={decorate, decoration={snake,amplitude=2.5pt}, draw},
	antivector/.style={decorate, decoration={snake,amplitude=-2.5pt}, draw},
    fermion/.style={draw=black, postaction={decorate},
        decoration={markings,mark=at position .55 with {\arrow[draw=black]{>}}}},
    fermionbar/.style={draw=black, postaction={decorate},
        decoration={markings,mark=at position .55 with {\arrow[draw=black]{<}}}},
    fermionnoarrow/.style={draw=black},
    gluon/.style={decorate, draw=black,
        decoration={coil,amplitude=4pt, segment length=5pt}},
    scalar/.style={dashed,draw=black, postaction={decorate},
        decoration={markings,mark=at position .55 with {\arrow[draw=black]{>}}}},
    scalarbar/.style={dashed,draw=black, postaction={decorate},
        decoration={markings,mark=at position .55 with {\arrow[draw=black]{<}}}},
    scalarnoarrow/.style={dashed,draw=black},
    electron/.style={draw=black, postaction={decorate},
        decoration={markings,mark=at position .55 with {\arrow[draw=black]{>}}}},
	bigvector/.style={decorate, decoration={snake,amplitude=4pt}, draw},
}

\tikzstyle{block} = [draw, rectangle, 
    minimum height=3em, minimum width=6em]

\usepackage{hyperref}			

\hypersetup{
    pdftitle={Long-Lived heavy stop at the LHC} 
   ,pdfauthor={rrr} 
   ,pdfsubject={}                        
   ,pdfcreator={}                        
   ,pdfproducer={}                       
   ,pdfkeywords={R-parity violation, Susy, LHC, CMS} 
   ,colorlinks=true                      
   ,linkcolor=black                      
   ,citecolor=black                      
   ,filecolor=black                      
   ,urlcolor=black                       
   ,pdfdisplaydoctitle=true              
   ,bookmarksnumbered=true               
}

\usetikzlibrary{shapes}
\usetikzlibrary{arrows}
\usetikzlibrary{positioning}

\title{
\vspace{-4.5ex}
{\normalsize \raggedright
}
\textbf{Long-Lived stop at the LHC with or without R-parity}
\vspace{2ex}
}

\author{L.~Covi and F.~Dradi \\[1ex]
{\normalsize\textit{Institut f\"{u}r Theoretische Physik, Friedrich-Hund-Platz 1, 
37077 G\"{o}ttingen, Germany}}
\vspace{3ex}
}
\date{}

\begin{document}

\maketitle

\thispagestyle{empty}

\begin{abstract}
\noindent
We consider scenarios of gravitino LSP and DM with stop NLSP both within R-parity 
conserving and R-parity violating supersymmetry (RPC and RPV SUSY, respectively). 
We discuss cosmological bounds from Big Bang Nucleosynthesis (BBN) and the gravitino 
abundance and then concentrate on the signals of long-lived stops at the LHC as
displaced vertices or metastable particles. 
Finally we discuss how to distinguish R-parity conserving and R-parity breaking 
stop decays if they happen within the detector and suppress SM backgrounds.

\end{abstract}

\newpage

\section{Introduction}
 
It is well known that cosmology provides evidence for physics beyond the Standard Model (SM), 
since in the SM there is no viable candidate for Dark Matter (DM). 
Within the Standard Model extensions providing a DM candidate, supersymmetry
is surely the most intensively studied, in particular the case of the lightest neutralino 
as a realization of the WIMP mechanism~\cite{SUSY-DM}.

Nevertheless supersymmetry provides also another possible DM candidate, 
the gravitino, which results from locally supersymmetric extensions of the 
standard model as gauge fermion of supergravity (SUGRA) \cite{Freedman:1976,Deser:1976}.
If it is not the LSP and not light, the gravitino causes grave cosmological problems 
since it can decay during or after primordial nucleosynthesis (BBN) and indeed strong 
constraints on the gravitino abundance and therefore the reheat temperature arise in
such scenarios~\cite{Gravitino-problem}.
One way to soften this problem is to require the gravitino to be the Lightest Supersymmetric 
Particle (LSP)~\cite{Bolz:1998} and in such case BBN constrains the possible late 
decays of the Next-to-Lightest Supersymmetric Particle (NLSP)~\cite{Kawasaki:(2008)}. 
In fact again the NLSP can have macroscopic lifetimes longer than one seconds and 
its decay or, if it is charged, even just its presence during BBN,  can change the 
primordial abundances of the light elements.

To avoid such constraints, it is sufficient to switch on a slight R-parity violation~\cite{Barbier:2004ez} 
and allow the NLSP to decay before BBN through R-parity violating couplings.
Even in  this case, the gravitino remains a viable Dark Matter candidate, since its decay
rate is doubly suppressed by the Planck mass and the small R-parity breaking parameter 
leading to a gravitino lifetime that exceeds the age of the universe by many orders of 
magnitude \cite{Takayama:2008,Buchmuller:2007ui}.  
Gravitino decays in the present epoch may lead to a diffuse $\gamma$-ray flux and 
comparing such signal with the diffuse $\gamma$-ray flux observed by Fermi-LAT 
collaboration~\cite{FermiLat1:2010,FermiLat2:2010},  a severe lower bound on 
the gravitino  lifetime of the order $ 10^{28} $ s, restricting the possible values
of the R-parity violation to $\epsilon \lesssim 2\times 10^{-8}$  \cite{Bobrovskyi:2010, Vertongen:2011mu}.
Such a value corresponds to  NLSP decay lengths above
$\mathcal{O}(\unit[50]{cm})$ - $\mathcal{O}(\unit[500]{m})$. 

We see therefore that in both gravitino DM scenarios with or without R-parity,
we expect the NLSP to have a long lifetime and give rise either to
displaced vertices or metastable tracks at the LHC. Note that even for average decay 
lengths larger that the size of the detector, a considerable fraction of the NLSP might still 
decay inside the detector and deserves dedicated researches from both a theoretical 
and experimental point of view~\cite{Ishiwata:2008tp,Graham:2012, Khachatryan:2011ts, Aad:2011yf, Chatrchyan:2012sp, Chatrchyan:2012dxa, 
Aad:2013pqd, CMS:collaboration_Long_Lived, Aad:2013gva, ATLAS_collaboration:2013}. 
In the last years a lot of attention has been given especially to the case of 
neutralino~\cite{Bobrovskyi:2010, Meade:2010ji, Bobrovskyi:2011vx, Hirsch:2005ag, Ghosh:2012pq, Covi:2010au},
stau~\cite{Steffen:2008bt, Endo:2010ya, Heisig:2011dr, Lindert:2011td, Heisig:2012zq, Heisig:2013rya}  or 
sneutrino~\cite{Covi:2007xj, Ellis:2008as, Katz:2009qx, Figy:2010hu, Roszkowski:2012nq} long-lived NLSP, 
since those states are more likely to be NLSP in the CMSSM or NUHM models. Recently also the case
of Higgsino NLSP was considered in \cite{Bobrovskyi:2012dc}.

In this paper, we investigate in particular stop NLSP decays, which have been previously
considered from the cosmological perspective in \cite{DiazCruz:2007fc, Berger:2008ti,  Kohri:2008cf}
and at the LHC as prompt decays in \cite{Alwall:2010jc, Kats-Shih:2011,Marshall:2014cwa}.  
The stop is the supersymmetric scalar with the largest left-right mixing and can therefore
naturally be the lightest colored state. Moreover it provides usually the largest
correction to the Higgs mass and its mass cannot be too large in order to
retain a light Higgs \cite{Asano:2010ut, Brust:2011tb, Papucci:2011wy}. 
Moreover the LHC constraints on the stops are much weaker than those on the gluino and 
first/second generation squarks since the stop production cross-section is smaller and 
its decay into tops or bottom more difficult to observe.
Our goal is the determination of the LHC reach for direct stop production for
long-lived stop NLSP, regardless of the decay channel and the comparison
of the parameter region with the cosmologically viable one.
Finally we will also discuss the possibility of distinguishing the RPC
and RPV decays if they happen in the LHC detectors, in particular if at least
one charged lepton is produced in the decay.

The paper is organized as follows: we discuss the general setting and the
stop-gravitino couplings in section 2 and then proceed to the cosmological
constraints from BBN and the DM density in section 3.
In section 4 we give our analysis of the LHC reach for the case of
displaced vertices in the pixel or tracker detectors and of metastable tracks.
Such analysis is independent on the decay channel, even if we will mostly 
assume that the decay product contain at least one lepton, e.g. a muon, that can be 
well-measured also away from the central part of the detector.
We will give the reach of LHC depending on the stop mass and lifetime
for both the different signals and see that there are interesting parameter
regions where two types of signals can be observed.
Finally in section 5 we will discuss how to distinguish the RPC and RPV
scenarios, assuming the decay happens in the detector.
We will then conclude in section 6.

\vspace*{0.5cm}

\section{Supersymmetric stop in the MSSM with and without R-parity}
\label{Supersymmetric stop NLSP}

We consider a supersymmetric model of the MSSM type given by the most general 
renormalizable superpotential. The R-parity, baryon and lepton conserving part
includes the Higgs $\mu$ term and Yukawa interactions as \cite{Martin:2011}
\begin{equation}
\label{B&LConservingSuperpotential}
W_0 = \mu H_u H_d + y^u_{ij} H_u Q_i U_j + y_{ij}^d H_d Q_i D_j + y_{ij}^e 
H_d L_i E_j\;,
\end{equation}
where the indices $i, j, k$ run over the three generations of fermions. 
The labels $L_i$, $E_i$, $Q_i$, $D_i$, $U_i$, $H_u$ and $H_d$ are for 
the chiral superfields containing the lepton doublet, lepton singlet, quark doublet, 
down-type quark singlet, up-type quark singlet, up-type Higgs doublet and 
down-type Higgs doublet, respectively. We denote with
$y_{ij}^{u}$, $y_{ij}^{d}$, $y_{ij}^{e}$ the dimensionless Yukawa 
coupling parameters, while $\mu$ is the supersymmetric higgsino mass parameter.

The baryon violating part of the superpotential is given by
\begin{equation}
\label{LViolationSuperpotential}
W_{\Delta B \ne 0} = \lambda_{ijk}^{''} U_i D_j D_k \;,
\end{equation}
while the lepton violating part is
\begin{equation}
W_{\Delta L \ne 0} = \lambda_{ijk} L_i L_j E_k + \lambda_{ijk}^{'} L_i Q_j D_k 
+ \mu_i L_i H_u\; ,
\end{equation}
where $\lambda_{ijk}$,  $\lambda_{ijk}^{'}$, $\lambda_{ijk}^{''}$ are 
R-parity violating Yukawa couplings and $\mu_i$ is a parameter with dimension 
of mass mixing the Higgs with the lepton multiplet.
If they all appear together, these parameters are strongly constrained by 
low-energy observable, i.e. proton decay, therefore in the following we
will consider only the case of either baryon violating or lepton
violating RPV.

Apart for the superpotential, we introduce in the lagrangian also the soft 
supersymmetry breaking terms, in particular gaugino and scalar masses
and soft trilinear terms $A_i$. Then the squared-mass matrix for the top 
squarks in the gauge-eigenstate basis $(\tilde t_L, \tilde t_R)$ is given by 
\begin{equation}
\label{stopmassmatrix}
\mathcal{L}_{m_{\tilde t}} = -\begin{pmatrix}\tilde t^*_R & \tilde 
t^*_L
\end{pmatrix}\,
\bold{m^2_{\tilde t}}\,\begin{pmatrix}\tilde{t}_L \cr \tilde{t}_R
\end{pmatrix}
\end{equation}
where 
\begin{equation}
\bold{m^{2}_{\tilde t}} = \begin{pmatrix}m_{\tilde{t}_R}^2 &  
m_t( A_t + \mu 
\cot\beta) \cr m_t (A_t + \mu\cot{\beta})  & m^2_{\tilde t_L} 
\end{pmatrix}
\end{equation}
is a non-diagonal hermitian matrix where $A_t$, $m_t$, and 
$\tan\beta$ denote, respectively, the trilinear coupling of the Higgs 
with top sfermions, the top quark mass, and the ratio of the two Higgs 
vacuum expectation values $ \tan\beta = v_u/v_d $.
 The masses $m_{\tilde t_R}^2$ and $m_{\tilde t_L}^2$ arise from the soft 
breaking,the D term contribution, and the top Yukawa coupling as follows:
\begin{eqnarray}
m^2_{\tilde t_R}&=& m^2_{\tilde U_3}+m^2_t + {2\over 3}\sin^2
{\theta_W} m^2_Z \cos 2{\beta} 
 \nonumber
\\ m^2_{\tilde t_L}&=& m^2_{\tilde Q_3}+m^2_t + ({1 \over 2}-{2 \over 3} 
\sin^2{\theta_W}) m_Z^2 \cos2{\beta}
\end{eqnarray}
where $\theta_W$ denotes the weak mixing angle and $m_Z$ is the $Z^0$ 
boson mass. The soft breaking masses $m_{\tilde U_3}$ and $m_{\tilde Q_3}$ 
are model-dependent. We see that in general the stop mass matrix can have
a large off-diagonal entry, in particular if $ A_t $ is chosen large to explain
the Higgs mass \cite{Hall:2011aa,Heinemeyer:2011aa, Arbey:2011ab}. 
In such a case the two mass eigenstate repel each other, so that the lightest one can 
become much lighter than the average mass scale.
The stop mass matrix $\bold{m^{2}_{\tilde t}}$ 
can be diagonalized by a unitary matrix to give mass eigenstates: 
\begin{equation}
\begin{pmatrix}
\tilde t_1\cr\tilde t_2 
\end{pmatrix} = 
\begin{pmatrix}\cos{\theta} & -\sin{\theta} \cr
          \sin{\theta} & \cos{\theta} \end{pmatrix}
\begin{pmatrix}\tilde t_L \cr \tilde t_R\end{pmatrix} .
\end{equation}
We define here $\tilde t_1 $ as the lightest state.

Note that different SUSY breaking scenarios can account for a stop NLSP with a
gravitino LSP. 
As an example we mention that in gauge-mediated 
supersymmetry breaking (GMSB) scenarios~\cite{Giudice:1998bp}
the supersymmetry-breaking scale is typically much smaller than in the gravity-mediated 
case, so that the gravitino is almost always the LSP. 
Moreover in the recently proposed model-independent 
framework of general gauge mediation (GGM)~\cite{Meade:2009,Buican:2009}
essentially any MSSM superpartner  can be the NLSP. 
Here we will assume that the NLSP is the lightest stop $\tilde t_1$.
At the present time many studies of light stops decaying promptly have been
performed,  see for instance \cite{Alwall:2010jc, Kats-Shih:2011} in the context of 
gravitino LSP, since the stop is the superpartner mostly connected to the Higgs
and required to be light to solve the hierarchy problem.
In this paper, we consider only non-prompt decays instead.
We take the stop masses as strongly split and we concentrate on the 
lightest stop mass eigenstate $\tilde t_1$, assuming that  
$\tilde t_2 $ and the rest of the colored supersymmetric particles
are outside the reach of LHC.
Note in any case that most of our results are very weakly dependent on the 
stop mixing angle and therefore valid also if the second stop is not too heavy,
as long as its production is suppressed. In case the other colored states
like the gluino and first two generation squarks are within the reach of
the LHC, we have additional particle production channels and the
search becomes more promising.

\newpage

\subsection{Stop NLSP couplings, production and decay channels}
\label{Stop coupling}

In this section we consider the interactions of the stop NLSP in 
supersymmetric models with gravitino LSP and DM candidate. 

The main interactions of the stop NLSP are the R-parity conserving QCD couplings, 
which in general dominate the stop pair production. In fact the R-parity violating couplings
considered here are many orders of magnitude smaller than the QCD gauge coupling and
too suppressed to give a measurable single-stop production.  
In the limiting case when the rest of the colored states are too heavy to be produced 
efficiently, the stop production cross-section is dominated by the direct production 
via the quark-antiquark annihilation and the gluon fusion channels. Then the stop 
mass is the only supersymmetric parameter influencing the production cross-section 
at tree-level and the dependence on the stop mixing arises only at NLO ~\cite{Beenakker:1997ut}.
In this paper we simulate the stop pair production at the LHC with  MADGRAPH 5~\cite{madgraph5:2011}, 
which includes only the LO cross-section and therefore we neglect any mixing angle dependence 
of the production. Note that NLO corrections can change the cross-section by a  of factor 
50-70\% within the mass range investigated here \cite{Beenakker:1997ut}. 
We will take such correction into account in our results by multiplying our cross-section 
by a constant NLO k-factor.

Let us now consider the decay channels of the stop and anti-stop pairs
in the RPC and RPV models.
In order to do that, let us recall that the gravitino comes into play when the supergravity 
is taken into account, i.e. supersymmetry is promoted to a local symmetry. In a sense, 
the gravitino plays the role of "gauge fermion" for local supersymmetry and its couplings
are fixed by supergravity. Therefore the coupling of stop with the gravitino is
 described by the SUGRA R-Parity Conserving lagrangian
\begin{equation}
 \label{int_lagr_of_grav}
 \mathcal{L}_{3/2} = -\frac{1}{M_P\sqrt{2}}\left[(\mathcal{D}_\nu
 \tilde t_R)^*\bar{\psi}_{3/2}^{\mu}\gamma^\nu \gamma_{\mu} P_R t 
 +  h.c.\right]
\end{equation}
where $\mathcal{D}_{\nu}\tilde t_R = \left( \partial_{\nu} + i g_i A_{\nu}^i
\right)\tilde t_R$ and $P_L(P_R)$ is the projection operator projecting 
onto left-handed (right-handed) spinor. Here, $A_{\nu}^i$ denotes any SM
gauge boson whereas $M_P = \left(8\pi G_N\right)^{-1/2}$ is the reduced 
Planck mass. The interaction lagrangian of the left-handed $\tilde t_L$ has 
an analogous form. We can easily obtain from this expression the
couplings for the lightest stop $\tilde t_1 $ as
\begin{equation}
 \label{int_lagr_of_grav-t1}
 \mathcal{L}_{3/2} = -\frac{1}{M_P\sqrt{2}}\left[(\mathcal{D}_\nu
 \tilde t_1)^*\bar{\psi}_{3/2}^{\mu}\gamma^\nu \gamma_{\mu} 
 (-\sin\theta P_R + \cos\theta P_L) t  + h. c. \right]
\end{equation}

In the rest frame of the decaying $\tilde t_i$, taking the matrix element 
at the order $ 1/m_{3/2}^2 $, the decay rate is independent 
of the mixing angle, which appears only in the interference at order 
$  m_t/m_{3/2} $,  and is given by the formula
\begin{equation}
 \label{stop_decay_rate_complete_formula}
 \Gamma_{\tilde t_1} = \frac{\left(m^2_{\tilde t_1}-m^2_{3/2}-m^2_{t}\right)^4}
 {48\pi m^2_{3/2}\,M_P^2 \,m^3_{\tilde t_1}}\left[1-\frac{4m_{3/2}^2 \,
 m_{\tilde t_1}^2}{\left(m^2_{\tilde t_1}-m_{3/2}^2-m_{t}^2\right)^2}\right]^{3/2},
\end{equation}
where $m_t = 173 \,\mbox{GeV}$ is the top mass, $m_{\tilde t}$ is the stop 
mass and $m_{3/2}$ is the gravitino mass. 

Now, if we neglect the top mass and the gravitino mass in the phase-space,
and normalize the stop mass to the value 
$500\,\mbox{GeV}$ and the gravitino mass to $1\,\mbox{GeV}$, we obtain a 
a stop lifetime given by 
\begin{equation}
 \label{lifetime_RPC_stop_decay}
\tau_{\tilde t_1} = \Gamma_{\tilde t_1}^{-1}\simeq (18.8\,\mbox{s}) 
\left(\frac{500 \,\mbox{GeV}}{m_{\tilde t_1}}\right)^5 \left(\frac{m_{3/2}}
{1 \,\mbox{GeV}} \right)^2 
\!\!\!\!.
\end{equation}
We see that the stop lifetime can cover a very large range of values since it strongly 
depends on the stop and gravitino masses. 
For instance, choosing $m_{\tilde t_1}=800\,\mbox{GeV}$, 
a lifetime of $\tau_{\tilde t_1}\simeq 0.018\,\mbox{s}$ or 
$\tau_{\tilde t_1}\simeq 1.9 \times 10^4\,\mbox{s}$ for a gravitino mass of 
$m_{3/2}=0.1\,\mbox{GeV}$ or $m_{3/2}=100\,\mbox{GeV}$ respectively, 
can be achieved. 
Finally, it is worth noting that this decay rate depends only on the particle masses. 

Let us now turn to the RPV couplings, in particular in the case of
bilinear RPV given in~\cite{Buchmuller:2007ui}. Other models with
unstable gravitino DM are given in \cite{Ji-Mohapatra:2008, Endo-Shindou:2009, Fileviez-Spinner:2009}.
In that case we assume that the only RPV coupling is
given by
\begin{equation}
\mu_i L_i H_u \subset W_{\Delta L \ne 0}\,, 
\end{equation}
which can be rotated away from the superpotential if we redefine $L_i$ and $H_d$ as 
\begin{equation}
\label{rotationLiHd}
L^{'}_{i} = L_{i} - \epsilon_{i} H_d\;,  \quad H_d^{'} = H_{d} + \epsilon_i 
L_i\; \quad\mbox{with}\quad\epsilon_i \equiv \frac{\mu_i}{\mu}\,.  
\end{equation}
However, in this new basis, R-parity breaking is reintroduced in the form of 
trilinear R-parity violation. So much so that, one obtains the new trilinear R-parity violating terms
\begin{equation}
\label{trilinearRPVtermsAfterRotation} 
\Delta W' = h_{ijk}L^{'}_i L^{'}_j E_k + h_{ijk}^{'} L^{'}_i Q_j D_k 
\end{equation}
where 
\begin{equation}
\label{definitionParametershTRPV} 
h_{ijk} = -y_{ij}^e \epsilon_k + y_{kj}^e \epsilon_i\;, \quad h^{'}_{ijk} 
= - y_{ij}^d \epsilon_{k}\,.
\end{equation}
and $\epsilon_i$ are the Bilinear R-Parity breaking parameters. 
We see therefore that the lightest stop can decay via this $ \lambda'$-type
coupling $h'$ through its LH component. The corresponding R-Parity violating 
lagrangian is in fact given by 
\begin{equation}
 \label{RPV_lagrangian_stop_decay}
 \mathcal{L} = \sqrt{2}\,\frac{m_b}{\upsilon_d}\,\epsilon_i \, \tilde t_L \bar{b}\,
  P_L \ell_i\, + h.c. =  \sqrt{2}\,\frac{m_b}{\upsilon\cos\beta}\,\epsilon_i\,\sin{\theta}\,
 \tilde t_1 \bar{b}\, P_L \ell_i\, + h.c.
\end{equation}
where the last expression is given in term of the stop mass eigenstate, the SM vacuum 
expectation value (vev) $\upsilon$ and the bottom mass.

In the rest frame of the decaying stop $\tilde t_1$, if the antilepton masses 
are neglected, the total decay rate for equal $\epsilon_i =\epsilon$ reads then
\begin{equation}
 \label{decay_rate_RPV_stop_Decay}
 \Gamma_{\tilde t_1} = \frac{3}{8\pi} \,
 \left(\frac{\epsilon\,\sin{\theta}\,m_{b}}{\upsilon\cos\beta}\right)^2\, m_{\tilde t_1} \,\left(1-\frac{m_b^2}
 {m_{\tilde t_1}^2}\right)^2, 
\end{equation} 
and so, using the explicit value for vev $\upsilon$ and bottom mass $m_b = 4 \,\mbox{GeV} $, 
the corresponding lifetime $\tau_{\tilde t_1}$ is 
\begin{equation}
 \label{lifetime_RPV_Stop_decay}
 \tau_{\tilde t_1} = 4.3 \times 10^{-7}\,\mbox{s} \left(\frac{\epsilon
 \sin{\theta}/\cos\beta}{10^{-8}}\right)^{-2}\!\!\left(\frac{500 \,
 \mbox{GeV}}{m_{\tilde t_1}}\right)^{-1}.
\end{equation}

Note that this stop decay is also present in the case of trilinear lepton number
RPV, while it is absent in the case of baryon number violating RPV.

We consider here this RPV stop decay as particularly promising because it
contains leptons in the final state, that can be more easily detected at 
collider experiments also away from the central collision region and are
therefore a very favorable signal.

In the case of baryonic violating RPV or even in MFV models like~\cite{Nikolidakis:2007fc, Csaki:2011ge}, 
the relevant superpotential coupling is given by the $ \lambda '' $ coupling and the lagrangian reads instead
\begin{equation}
 \label{B-RPV_lagrangian_stop_decay}
 \mathcal{L} = 
 \sqrt{2}\, \lambda''_{3jk} \tilde t_R \bar d_j P_L d_k^c + h.c. 
 =  \sqrt{2}\, \lambda''_{3jk} \cos\theta\; \tilde t_1 \bar d_j P_L d_k^c + h.c. \; ,
\end{equation}
with $ \lambda_{3jk}'' $ antisymmetric on the last two indices, 
giving the decay rate into two light-quark jets as
\begin{equation}
 \label{decay_rate_BRPV_stop_Decay}
 \Gamma_{\tilde t_1} =  \left( \lambda''_{321} \cos\theta \right)^2\;
 \frac{ m_{\tilde t_1}   }{8\pi}   \; .
\end{equation} 
For small $ \lambda'' $ this decay can also lead to displaced vertices, with
two jets originating far away from the stop pair production vertex.

In any of the scenarios we discussed here, we see that the stop lifetime is always 
much longer than the hadronization time and therefore we expect the stop and anti-stop
to hadronize into an R-hadron before decay~\cite{Gates:1999ei}. 
Such an R-hadron can in principle be both electromagnetically charged or not and 
even change its charge while it travels in the detector \cite{Fairbairn:2006gg}.

For the sake of clarity and transparency of exposition, from now on we adopt 
the convention that the lightest stop $\tilde t_1$ is simply called $\tilde t$. 

\vspace*{0.5cm}

\section{Stop NLSP and gravitino LSP in cosmology}

In this section we discuss very shortly the cosmological bounds
on the scenario with stop NLSP and the gravitino DM and LSP, in order 
to single out the cosmologically preferred parameter space in both RPC and RPV models.

Let us consider first the effect of a stop NLSP during BBN.
The stop is a colored and EM-charged particle and therefore
it can disrupt BBN not only through the energy release in the
decay, but also because of bound state effects~\cite{Jedamzik:2009uy}.
In the first case the light element abundances are more strongly
affected by hadro-dissociation and therefore the limits are
more stringent for hadronically decaying particles like the stop~\cite{Kawasaki:2004qu, Jedamzik:2006xz}.
In the latter case the constraints are independent of the decay channel and just depend 
on the stop lifetime and density at the time of decay  \cite{Jedamzik:2009uy, Kusakabe:2009}.   
They therefore apply equally to any of the scenarios we discussed.

The limits on the abundance $Y_{\chi}(\tau_\chi)= n_{\chi}/s$ from bound
state effects for a hypothetical long-lived strongly interacting massive particle $\chi$, 
have been computed by M. Kusakabe {\it et al.} in \cite{Kusakabe:2009}. 
Requiring that the primordial light element abundances remain within
the observed ranges, they obtained the following constraints
depending on the particle lifetime $ \tau_\chi $:
\begin{itemize}
\item $Y_\chi < 10^{-18}-10^{-12}$ (for $30$~s $< \tau_\chi < 200-300$~s), 
\item $Y_\chi <10^{-18}-10^{-21}$ (for $200-300$~s $< \tau_\chi < 2 \times 
10^3$~s) \item $Y_\chi < 10^{-21}-10^{-22.6}$ (for 2$\times10^3$~s $< 
\tau_\chi \ll 4\times 10^{17}$~s)\; .
\end{itemize}
In the window ($30~\mbox{s} < \tau_\chi < 200-300$~s)
the most stringent constraint comes from the upper limit on the $^7$Li, 
while for  ($200-300~\mbox{s} < \tau_\chi < 2 \times 10^3$~s) the strongest 
constraint is due to the upper limit on the B abundance.
Finally the upper limit on the $^9$Be abundance determines the
bound for longer lifetimes.
These constraints are very steep and become quickly dominant over the 
hadro-dissociation bounds ~\cite{Kawasaki:2004qu, Jedamzik:2006xz}. 
For lifetimes shorter that  approximately 30 s  the constraints from bound states
disappear and those on hadronic decays are very weak~\cite{Kohri:2001jx}
and this corresponds in our two scenarios to a stop mass range
\begin{equation}
 \label{lifetime_noBBN}
 m_{\tilde t} \geq \left\{
 \begin{array}{cc}
451\; \mbox{GeV} \left(\frac{m_{3/2}}{1\, {\rm GeV}} \right)^{2/5}  & \mbox{RPC} \cr
1.4 \times 10^{-8}\; \mbox{GeV} \left(\frac{\epsilon\sin{\theta}/\cos\beta}{10^{-8}}\right)^{-2} & \mbox{RPV}
\end{array} \right. .
\end{equation}
We see therefore that BBN does not provide practically any bound on the RPV scenario,
apart if the RPV coupling is very small, below $10^{-12} $. 
We will therefore in the following only consider in detail the constraints for the RPC case.

\subsection{RPC decay of stop NLSP in cosmology}
\label{cosmological_analysis_NLSPStop}

Assuming that the stop NLSP is in equilibrium in the thermal plasma,
the density of the stop at chemical decoupling is given by the
solution of a Boltzmann equation as in the case of a WIMP:
\begin{equation}
 \label{boltzmann_equation}
 \frac{dY_{\tilde t}}{dx}= -\frac{x s(x)}{H(x) m^2_{\tilde t}}\left\langle
 \sigma_{ann} v \right \rangle_{x}\left(Y_{\tilde t}^2 -Y_{eq}^2 \right),
\end{equation}
where $x=m_{\tilde t}/T$, $H$ is the Hubble parameter,
$Y_{eq}$ is the equilibrium stop abundance and $\left\langle\sigma_{ann} 
v \right \rangle_{x}$ is the thermally averaged annihilation cross section. 

The density of a colored relic and the BBN bounds have been studied in a model 
independent way by C. Berger {\it et al.} in \cite{Berger:2008ti} . We follow here their analysis
just updating the constraints to those given above.
In \cite{Berger:2008ti} the authors first considered the simplified case of a single
annihilation channel $\tilde t \tilde t^*\to gg$. Such a choice was motivated 
by the fact that this channel just depends on the stop mass and its QCD representation, 
without dependence on the rest of the supersymmetric spectrum, and, in addition, 
it is always the dominant channel, contributing at least $50\%$ of the
total annihilation cross-section.
This channel gives the most conservative result since it cannot be suppressed
by particular choices of spectrum or mixing and it provides a conservative
upper limit on the stop abundance. In fact, if  more annihilation channels are taken 
into account, the cross section increases and, therefore, the stop density decreases. 
In \cite{Berger:2008ti} also the computation of the Sommerfeld enhancement~\cite{Sommerfeld} 
for this particular channel was performed with two different prescriptions for the higher orders.
The Sommerfeld enhancement increases the cross-section at low velocity
and can be obtained by resumming over the exchange of a ladder of gauge bosons
between the initial particles.
 It was found that the averaged Sommerfeld factor reduces the tree-level yield 
 by roughly a factor of $2$, while the summed  Sommerfeld one by roughly a 
 factor of $3$.
 We will therefore take the stop abundance from the leading order 
 computation and vary it by a factor 2-3 to see the effect of both the
 Sommerfeld enhancement and the additional annihilation channels.
  
The stop abundance is proportional to the stop mass and it 
can be rescaled as
\begin{equation}
 \label{rescaled_abundance_with_mass}
 Y(m_{\tilde t})= Y(1\,\mbox{TeV}) \left(\frac{m_{\tilde t}}{1\,\mbox{TeV}}
 \right)\end{equation}
up to logarithmic corrections, since, in general, the mass always enters linearly 
in the coefficient of Eq.(\ref{boltzmann_equation})~\footnote{
Recently it has been discussed in \cite{Heisig:2013rya} that the stau NLSP abundance
is better fitted by a dependence given by $ m_{\tilde\tau}^{0.9} $ but in the
range of masses we are considering such a difference in the exponent has 
negligible effect.}. 

\begin{figure}[t]
\centering
\includegraphics[scale=0.90]{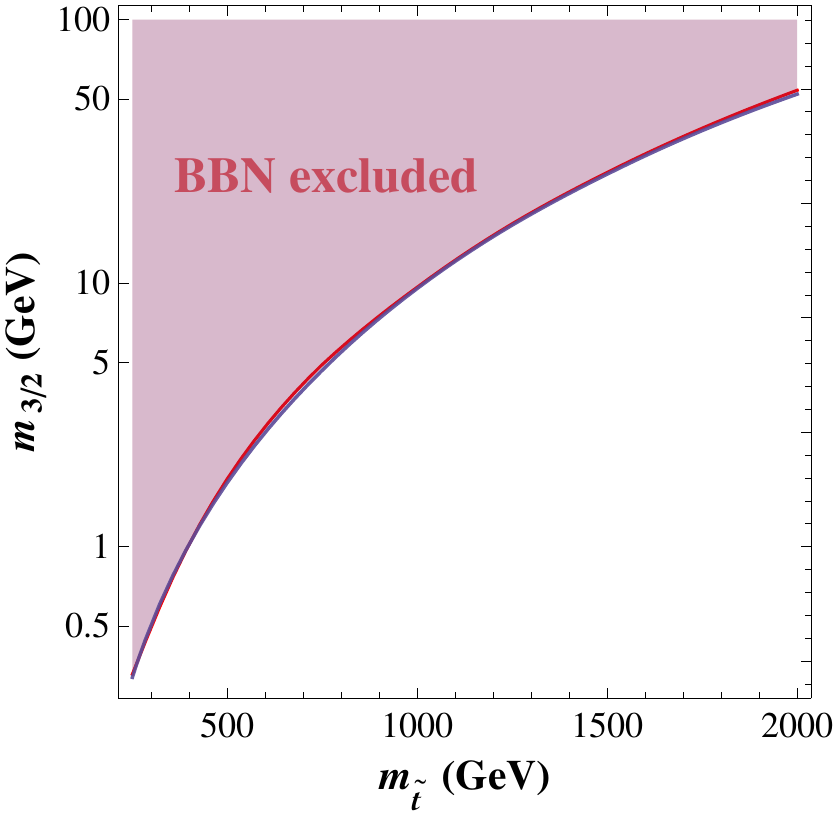}
\caption{BBN bounds on the PRC stop NLSP with gravitino LSP scenario.
The red and (superimposed) blue regions are excluded by the BBN constraints
assuming the LO or the averaged Sommerfeld enhanced stop abundance.}
\label{fig:BBNbound}
\end{figure}

In order to set limits on the RPC model, we first compute the stop density
as a function of the stop mass from Eq.~(\ref{rescaled_abundance_with_mass})
and we compare it with the limits in \cite{Kusakabe:2009}.
We determine then the maximal allowed value of the stop lifetime for
the particular mass by the value of $ \tau_{\tilde t_1}$ such that 
$ Y_{\tilde t_1} = Y_\chi^{bound} $. Through the analytical formula
for the stop lifetime, we can then recast the bounds in the plane
$ m_{\tilde t} $ vs $ m_{3/2} $. 
We give these results in Fig.~\ref{fig:BBNbound}.
We see that the constraints for the LO and the Sommerfeld enhanced case
overlap and are almost in perfect agreement. This is due to the fact that the
bound-state BBN constraints are very steep and do not change appreciably
even if the stop density changes by a factor of a few.
So the BBN bounds practically do not depend on the details of the
stop freeze-out, as long as no strong resonant annihilation is present, 
and are quite robust and independent of the masses of the
heavier superpartner and therefore of the particular supersymmetric model
with stop NLSP~\footnote{We do not consider here the case of coannihilations for
the stop, but note that coannihilation with a more weakly-interacting state
like the neutralinos/sleptons increases the number density and therefore 
would correspond to stronger constraints.}.

\subsection{CDM constraints}

We are assuming in this paper that gravitinos are Cold Dark Matter
and therefore they must have obtained the required abundance in the
course of the cosmological evolution. The gravitino production by the
decay of the stop NLSP~\cite{Covi:1999,Feng:2003}, is in most of the 
parameter space negligible since either the stop abundance or the 
gravitino mass are too small. 
Moreover such contribution is substantial only in the case of the RPC
model and it is instead very much suppressed if the stop RPV decay
is dominant. 

On the other hand, gravitinos can be produced in substantial numbers by 
scatterings and decays of supersymmetric particles in equilibrium in the hot plasma. 
Their abundance is then proportional to the bath reheating temperature $ T_R$ 
and can exceed the critical density of the universe if no restrictions on the 
reheating temperature is imposed \cite{Moroi-Murayama:1993, Bolz:2007,Pradler:2006qh}. 
In this context the gravitino abundance $\Omega_{3/2}$ is given 
by~\cite{Bolz:2007,Pradler:2006qh, Covi:2010au}
\begin{equation}
\label{gravitino_density_formula}
\Omega_{3/2}\,h^2 \approx \frac{T_R}{10^9\,\mbox{GeV}}\left(\frac{m_{\tilde t}}
{300\,\mbox{GeV}}
\right)^2 \left(\frac{m_{3/2}}{1\,\mbox{GeV}}\right)^{-1}\sum_{i=1}^3 \gamma_{i}
\left(\frac{M_{i}}{m_{\tilde t}}\right)^2 
\end{equation}
where $M_i$ are the physical gaugino masses and the coefficients $\gamma_i$ 
account for RGE effects between the reheating temperature scale and 
the scale of the physical gaugino masses. We have for those constants
the ranges  $\gamma_1=  0.17-0.22$, $\gamma_2= 0.54-0.57$, 
$\gamma_3=0.48-0.52$ from the  1-loop RGE for the gaugino 
masses and gauge couplings from $T_R=10^7-10^9\,\mbox{GeV}$~\cite{Covi:2010au}.
We neglect here possible decay of the heavier superpartners in equilibrium, 
the FIMP  contribution~\cite{Hall:2009bx}, which may even play a dominant role in the case 
of hierarchical spectra between gauginos and scalar superpartners \cite{Cheung:2011nn}.
Note here the dependence on other supersymmetric masses than the LSP mass, 
such that the exact abundance becomes a model dependent quantity.
Since we are here interested mostly on a constraint on the gravitino and stop
mass, we will here consider as most conservative the case when the gaugino
masses are not much heavier that the stop, 
$ M_{i}/m_{\tilde t} = (1.1 -2) $, in order to minimize the gravitino production.

\begin{figure}[t]
\begin{center}
$\begin{array}{cc}
\hspace{-3mm}\includegraphics[scale=1]{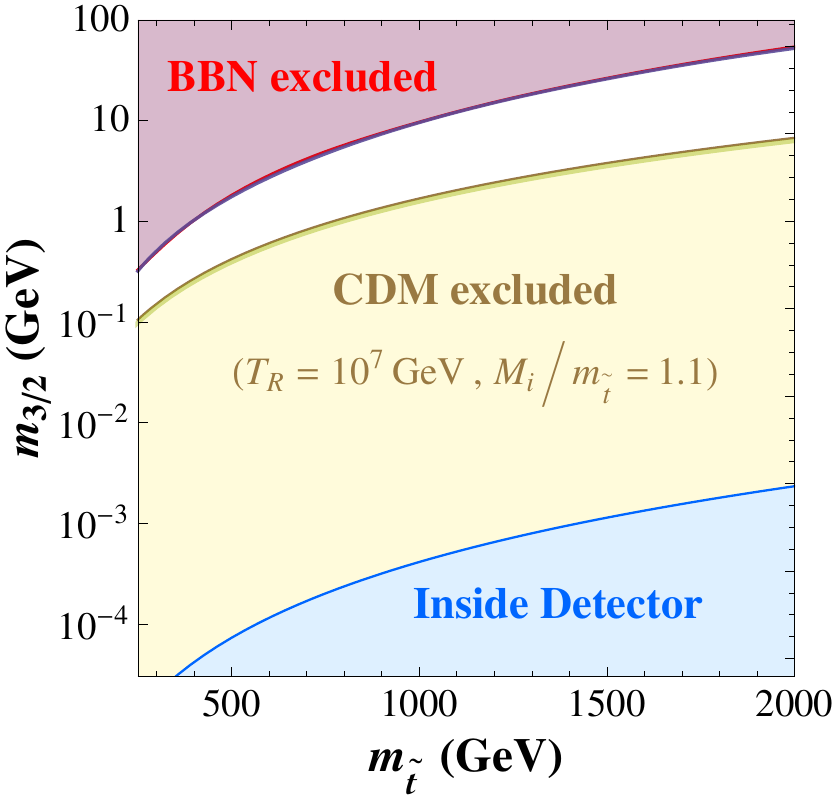}&
\includegraphics[scale=1]{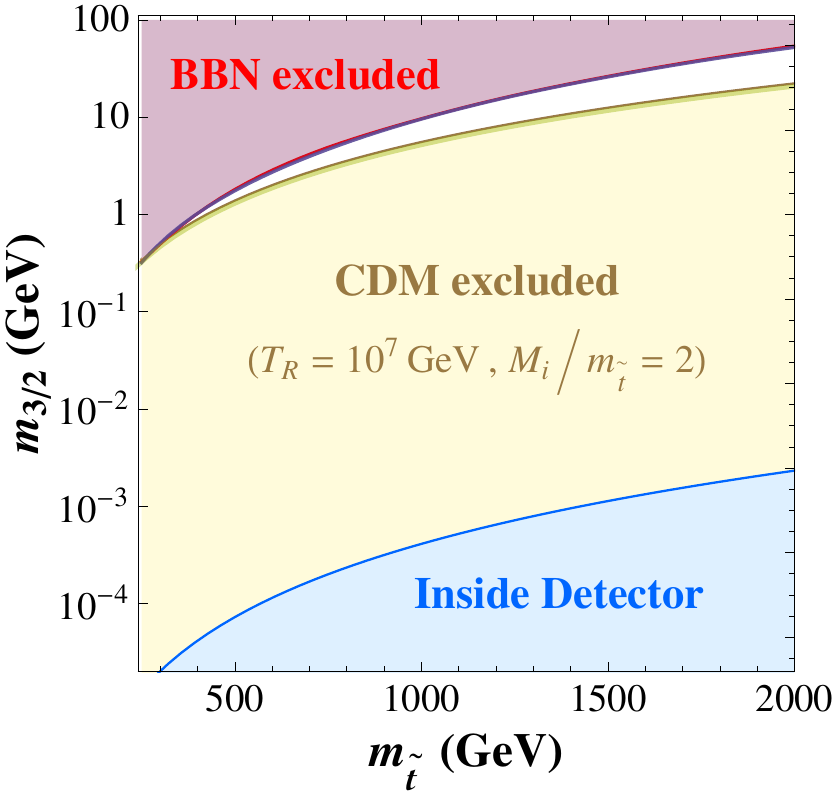}
\end{array}$
\end{center}
\vspace{-4mm}\caption{Plane "$m_{\tilde t}\;\mbox{vs}\;m_{3/2}$\!\!" with 
the BBN and CDM excluded regions (red and yellow, respectively) and with the 
Inside detector region (light blue). On the left side (right side) of the 
figure a CDM excluded region for $T_R=10^7\,\mbox{TeV}$ ($10^7\,\mbox{TeV}$) 
and $M_i/m_{\tilde t}=1.1$ (\,$2$\,) is drawn. The allowed region is painted 
white.}
\label{fig:Allbounds7112}
\end{figure}

The current best-fit values for Dark Matter density in the universe from the data
of the Planck satellite is given by~\cite{Ade:2013zuv}
\begin{equation}
 \label{best_fit_OmegaDM-Planck}
 \Omega_{\mbox{\tiny{CDM}}}h^2=0.1199(27),
\end{equation}
and results slightly larger that the previously obtained combination of the seven-years 
WMAP data, observations of baryon acoustic oscillations and determinations of the 
present Hubble parameter in
\cite{Komatsu:2011} ,
\begin{equation}
 \label{best_fit_OmegaDM}
 \Omega_{\mbox{\tiny{CDM}}}h^2=0.1126(36).
\end{equation}
We impose here that the gravitino energy density in Eq.(\ref{gravitino_density_formula})
is smaller or equal to the the Cold Dark Matter density. From the equality to two 
measurements $\Omega_{\mbox{\tiny{CDM}}} = 0.1126-0.1199 $ 
 we obtain the yellow and brown curves in 
 Fig.~\ref{fig:Allbounds7112}. 
On such lines the gravitino CDM density is fully produced by thermal scatterings,
while below the line the gravitinos are overabundant and therefore the parameter
space is excluded by the CDM constraint.
Note that the position of the line depends on the particular $T_R$ assumed
and the curves move up and down in the value of the gravitino mass exactly
by the change in $T_R$, since indeed the dependence is on $ T_R/m_{3/2}$.
Finally, comparing the two plots for different values of 
$ M_i/m_{\tilde t} $, we can see that for a bigger ratio of physical gaugino masses 
to stop mass, the white allowed region is reduced.

\subsection{All constraints}
\label{subsection:allconstraints}

We plot now both the cosmological constraints together in the plane
$m_{\tilde t}\;\mbox{vs}\;m_{3/2}$ in  Fig.~\ref{fig:Allbounds7112}. 
For future convenience we also show the region of the parameter
space where the RPC stop lifetime is smaller than  $10^{-7} $
and therefore the stop decays inside the detector.

Looking at this figure, we see that the allowed region is limited 
from above by the BBN constraint and from below by the CDM
constraint, so that only a narrow allowed strip remains for $T_R =10^7\,\mbox{TeV}$.
The breadth of such strip depends both on the supersymmetric
spectra and in particular on $ M_i/m_{\tilde t} $ and on the
reheat temperature assumed. In particular for reheat
temperatures above a few $10^{7}$ GeV no allowed parameter
space remains for stop masses below 2 TeV.

We note that the region where the RPC decay is sufficiently
fast to happen in the detector correspond to a very low reheat
temperature of the order of $ 10^3-10^4 $ GeV, so that 
in case of high reheating temperature the stops appear 
as metastable particles at the LHC.
\begin{figure}[!ht]
\begin{center}
$\begin{array}{cc}
\hspace{-3mm}\includegraphics[scale=0.9]{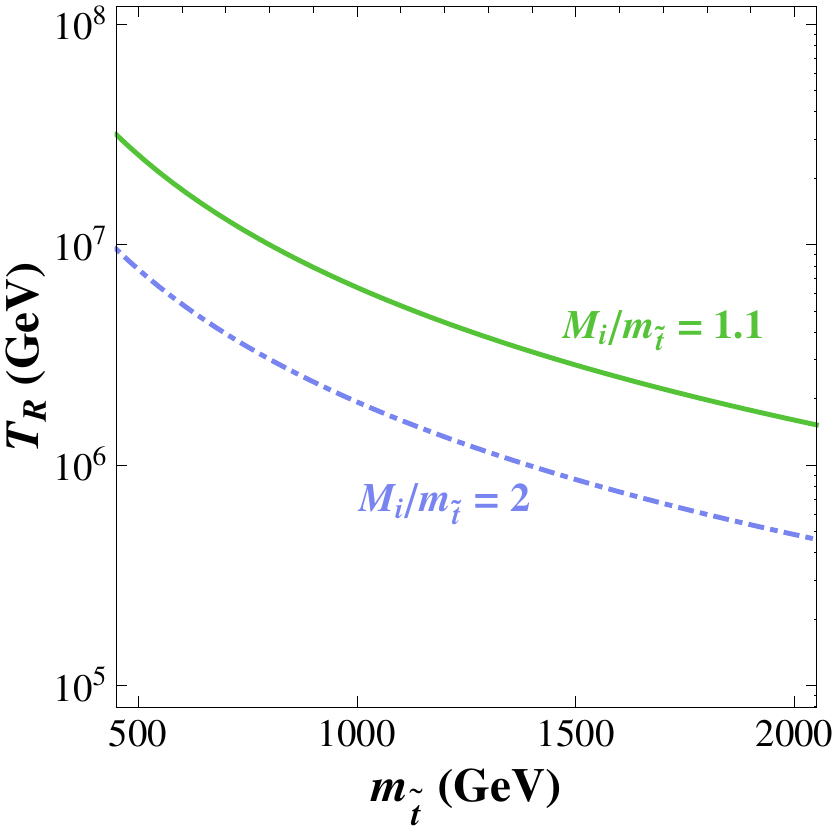}&
\includegraphics[scale=0.95]{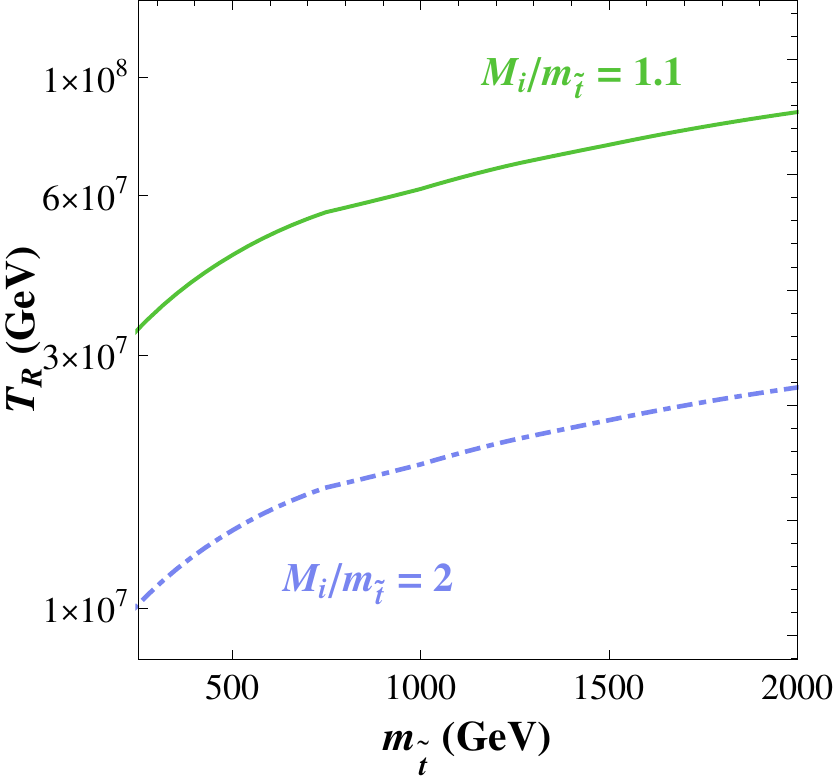}
\end{array}$
\end{center}
\caption{Plot of the reheating temperture $T_R$ as a function of the 
stop mass $m_{\tilde t}$ for the right value of DM density and $m_{3/2}=1\mbox{GeV}$ 
on the left panel. 
Plot of the maximal value of $T_R$ as a 
function of the stop mass $m_{\tilde t}$ on the right panel. In both 
plots the curves for the ratio $M_i/m_{\tilde t}=1.1$ (green solid line) 
and $M_i/m_{\tilde t}=2$ (blue dot-dashed line) are plotted.}
\label{fig:TrStopmass}
\end{figure}

In order to conclude the discussion of the cosmology of a 
gravitino CDM with stop NLSP, it is worth drawing the reheating temperature 
corresponding to the right value of Dark Matter density for
$ m_{3/2} =  1$ GeV  and the maximal allowed reheat temperature $ T^{max}_R $
as a function of the stop mass in n Fig.~\ref{fig:TrStopmass}.
Such curves are shown for two different values of the ratio of physical gaugino 
masses and  stop mass in Fig.~\ref{fig:TrStopmass}, e.g. $
M_i/m_{\tilde t}=1.1$ in green (solid line) and $M_i/m_{\tilde t}=2$ in blue 
(dot-dashed line). We see that for larger NLSP mass, a smaller reheating temperature 
is needed to match the observed DM density at fixed gravitino mass, while on 
the other hand the bound on $ T_R$ becomes relaxed for larger stop masses as 
the BBN bounds are weaker.

\newpage

\section{Decay of stop NLSP at LHC}

Many extensions of the SM include heavy, long-lived, charged particles 
(HSCPs). These  particles can travel distances comparable to the size 
of modern detectors, where they might be produced. Thus, they might 
appear to be stable if they have a lifetime bigger that a few nanoseconds. 
Moreover, the HSCPs can be singly charged ($|Q| = 1e$), fractionally 
charged ($|Q| < 1e$), or multiply charged ($|Q| > 1e$). Since the particle 
identification algorithms at hadron collider experiments generally assume 
signatures appropriate for SM particles, e.g., $v \approx c$ and $Q = 0$ 
or $\pm 1e$, nowdays the HSCPs might be misidentified or even completely 
missed without dedicated searches. 
The LHC experiments have already performed specific analysis, especially
for the case of metastable particles 
\cite{Khachatryan:2011ts, Aad:2011yf, Chatrchyan:2012sp, Chatrchyan:2012dxa, 
Aad:2013pqd, CMS:collaboration_Long_Lived, Aad:2013gva}.

In this very exciting background the goal of this section is to study two 
different classes of signal coming from a long-lived stop NLSP, which is 
produced by the proton-proton collision at LHC, at a centre of mass energy 
of $\sqrt{s}=14\,\mbox{TeV}$ and an integrated luminosities of $L=25\,
\mbox{fb}^{-1}$ and $3000\,\mbox{fb}^{-1}$. 
The first signal is represented by a displaced vertex inside the detector due 
to the stop decay inside the Pixel or Tracker detector. We will here mostly 
consider the kinematics and geometry of the CMS detector to estimate the 
number of decaying events within two adjacent detector parts. We neglect 
the interactions of the R-hadron with the detector material that could cause 
the particle to stop in the detector before the decay and the presence of a magnetic 
field bending the trajectory for charged R-hadron. Such effects could be taken
into account only by a full detector simulation, which is beyond the scope
of our study. Note in any case that interactions in the detector can only lead 
to a larger number of events inside the detector and that the magnetic field
does not affect neutral R-mesons, which are expected to be around 50\% of the 
cases~\cite{Gates:1999ei}.

The second type of signal is instead the HSCP track of a metastable stop
that leaves the detector before decaying. Such a signal is actively searched
for by the LHC collaborations.

In both cases we will use MadGraph 5 to compute the LO stop production at 
the LHC and we will correct it with a constant NLO k-factor of $1.6$ corresponding 
to the k-factor given by Prospino~\cite{Beenakker:1996ed} for a stop mass of $800$ GeV. 
We checked that this factor remains in the range $1.5-1.7$ for stop masses up to 2 TeV.

Regarding the decay, we will either include it within the MadGraph 5
analysis with a reference decay rate and then rescale the distances to
probe the whole accessible lifetime range or use an analytical estimate for the 
distribution of the decay lengths. As we will see both approaches give similar 
results, with the semi-analytical one allowing to explore more easily the parameter 
space.
Before going into detail, we want to highlight here that this analysis is 
independent from the stop decay channel, as long as the decay
gives measurable tracks and a clear displaced vertex. 
In fact, we will apply our results to the parameter space of both RPC and 
RPV models that we have already discussed in Sec.~\ref{Supersymmetric stop NLSP}. 
We are returning to this crucial point later in 
Sec.~\ref{Detector analysis of both RPC and RPV}.

To this day, both the CMS and ATLAS experiments have published searches 
on this topic for HSCPs produced in proton-proton collisions. Their latest 
results can be found, respectively, in the papers 
\cite{CMS:collaboration_Long_Lived, Aad:2013gva}. In particular, 
the CMS analysis investigates signatures in three different parts of the 
detector using data recorded at a centre of mass energy of $\sqrt{s}=7\,
\mbox{TeV}$ and $8\,\mbox{TeV}$ and an integrated luminosity of $L=5\,
\mbox{fb}^{-1}$ and $18.8\,\mbox{fb}^{-1}$. The studied parts are the inner 
tracker only, the inner tracker and muon detector, and the muon detector 
only.
On the other hand, the ATLAS collaboration is also investigating signatures producing 
a displaced multi-track vertex with at least a high transverse momentum muon at a distance 
between millimeters and tens of centimeters from the proton-proton interaction point, 
by using a data sample at $\sqrt{s} = 8\,\mbox{TeV}$ and for $L=20.3\,\mbox{fb}^{-1}$ \cite{ATLAS_collaboration:2013}. 

As a detector, in this paper we will consider as explicit example the CMS experiment 
even if we expect ATLAS to have comparable reach, perhaps even larger due
to the bigger size. So first of all we start describing 
very briefly how the CMS detector is made and what it searches for. 

\subsection{The CMS detector}

The CMS detector uses a right-handed coordinate system where the 
origin is at the nominal interaction point. The $x$-axis points 
towards the centre of the LHC ring, the $y$-axis points up with 
respect to the plane of the LHC ring and, at last, the $z$-axis along 
the counterclockwise beam direction. The polar angle $\theta$ 
is measured from the positive $z$-axis, the azimuthal angle $\phi$ in the 
$x\!-\!y$ plane and the radial coordinate in this plane is denoted by $r$. 
The transverse quantities, such as the transverse momentum ($\vec{p_T}$), 
always refer to the components in the $x\!-\!y$ plane. In this context, 
the magnitude of the three-vector $\vec{p_T}$ is indicated by $p_T$ and 
the transverse energy $E_T$ is defined as $E\sin{\theta}$.

\begin{figure}[t]
\centering
\includegraphics[scale=0.5]{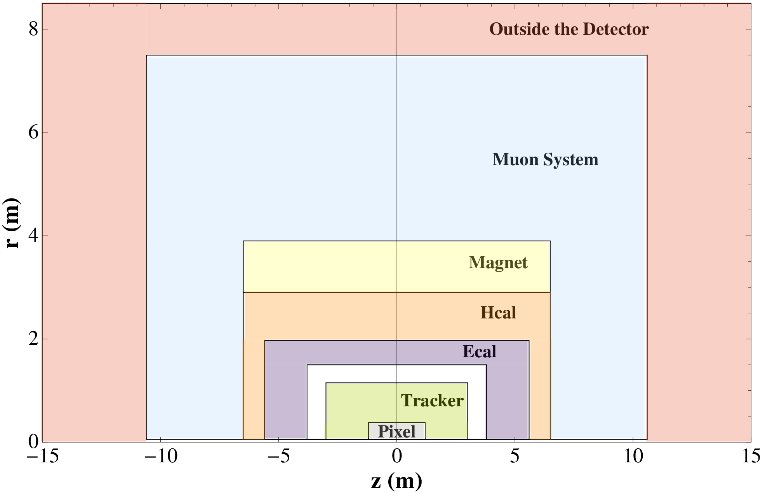}
\caption{Layout of two quarters of CMS detector used in this analysis 
similar to \cite{Bobrovskyi:2011vx}}
\label{fig:longitudinalView}
\end{figure}

In order to make easier the description of the CMS detector, the layout 
of one quarter of it was sketched in Fig.~\ref{fig:longitudinalView}. Now, 
if we start from the innermost part of the detector and going outwards, 
we can see the following parts: 
\textit{Interaction Point (IP)}, \textit{Pixel (Pi)}, \textit{Tracker (Tr)}, 
\textit{Electromagnetic Calorimeter (EC)}, \textit{Hadron Calorimeter (HC)}, 
\textit{Magnet (M)}, \textit{Muon System (MS)}. 
Below, a very short description of all of them is listed. A more detailed 
one, instead, can be found in \cite{CMS:2008}.  
\begin{itemize}
  \item \textit{Interaction Point (IP)} is the point in the centre of 
  the detector at which proton-proton collisions occur between the two 
  counter-rotating beams. We will assume that the stop and anti-stop pair
  is produced at this point. 
  
 \item \textit{Pixel (Pi)} detector contains 65 million pixels, 
 allowing it to track the paths of particles emerging from the collision 
 with extreme accuracy. It is also the closest detector to the beam pipe 
 and, therefore, is vital in reconstructing the tracks of very short-lived 
 particles. We therefore expect that the decay would be very well
 measured if it happens in this part of the detector. 
 
 \item \textit{Tracker (Tr)} can reconstruct the paths of high-energy muons, 
 electrons and hadrons, as well as see tracks coming from the decay of very 
 short-lived particles. It is also the second inner most layer and, so, receives 
 (along with the Pixel) the highest number of particles.
 Even if it is less densely equipped than the Pixel detector, it can still
 recognize tracks coming from a displaced vertex instead than the interaction point.
 
 \item \textit{Electromagnetic Calorimeter (EC)} is designed to measure the 
 energies of electrons and photons with high accuracy via electromagnetic 
 calorimeters. In our case it can allow to measure the energy of the 
 lepton arising in the decay.
 
 \item \textit{Hadron Calorimeter (HC)} measures the energy of hadrons
 and can give an estimate of the b-jet energy in the decay. 
 
 \item \textit{Magnet (M)} is the central device around which the experiment 
 is built.
 The job of this big magnet  ($\vec{B}=4$ T), which contains all the parts 
 above, is to bend the paths of particles and allow for an accurate 
 measurement of the momentum of even high-energy particles. 
 
 \item \textit{Muon System (MS)} is able to detect muons and possibly
 other charged particles able to cross the whole detector. 
 \end{itemize}

The best detector parts to single out the presence of a displaced vertex are
the pixel and tracker detectors and therefore we will restrict our discussion to
the case of stop/anti-stop decaying there or surviving through the whole detector.

\subsection{Numerical analysis}

The numerical analysis in this paper is realized by means of the open source 
software MadGraph~5 which can generate matrix elements at tree-level, given 
a lagrangian based model, for the simulation of parton-level events for decay 
and collision processes at high energy colliders \cite{madgraph5:2011}. 
In order to study the stop production at the LHC experiment we choose the 
MSSM model from the MadGraph 5 library of models which is built upon that 
of the package FeynRules\footnote{FeynRules is a Mathematica-based package 
which addresses the implementation of particle physics models, which are 
given in the form of a list of fields, parameters and a Lagrangian, into 
high-energy physics tools. It calculates the underlying Feynman rules and 
outputs them to a form appropriate for various programs such as CalcHEP, 
FeynArts, MadGraph, Sherpa and Whizard.}\cite{feynrules:2013}. Furthermore, 
we choose that the proton-proton colliding beams are not polarized and the 
centre of mass energy is $\sqrt{s}=14 \;\mbox{TeV}$ for all our simulations. 

\begin{figure}[!ht]
\centering
\includegraphics[scale=0.7]{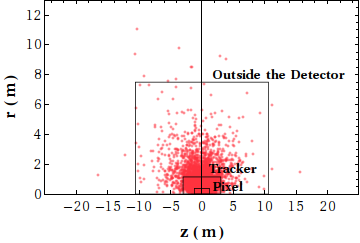}
\caption{Distribution of $10,000$ stop displaced vertices in the 
detector plane $(r,z)$ for $m_{\tilde t}=800\,\mbox{GeV}$ and 
$\Gamma=2.02159\times 10^{-16}\,\mbox{GeV}$ along with the size 
of Pixel, Tracker and the whole CMS detector.}
\label{fig:distribution_on_Rz-plane}
\end{figure}

We run MadGraph 5 for several stop masses but for only one reference value of
the stop decay rate that we take to be $\Gamma_{\tilde t}^{cm}=2.02159
\times10^{-10}\,\mbox{GeV}$, requiring that it has to generate $10,000$ 
events per run. By doing so, the kinematics of all particles in the 
process can be obtained. 
From those quantity we compute numerically the stop and anti-stop decay lengths 
$\ell_{\tilde t}$ in each single event assuming that the particles propagate undisturbed 
from the interaction point and decay randomly according to an exponential distribution.
From the decay length and the stop or anti-stop momentum direction we can 
draw using the software Mathematica the distribution of all decay vertices 
in the detector plane $(r,z)$. 
If we note that any change in the stop decay rate can be compensated 
by a change in the distance that the stop travels, we can circumvent the 
problem of launching MadGraph 5 for all of decay rates we need for our 
analysis, by simply rescaling the dimensions of all parts of the detector 
consistently. This point will be clarified in the next subsection when we 
are discussing the semi-analytic analysis. By using this expedient, the 
stop length distribution for a stop mass of $m_{\tilde t}=800\,\mbox{GeV}$ 
and a stop decay rate of $\Gamma_{\tilde t}=2.0159\times10^{-16}\,
\mbox{GeV}$ can be easily found from the reference data by rescaling
the size of the detector by a factor of $10^{-6}$. Such a distribution is shown 
in Fig.~\ref{fig:distribution_on_Rz-plane} by red dots along with the size of 
pixel, tracker and the whole CMS detector. 

\begin{figure}[t]
\begin{center}
$\begin{array}{cc}
\hspace{-3mm}\includegraphics[scale=.90]{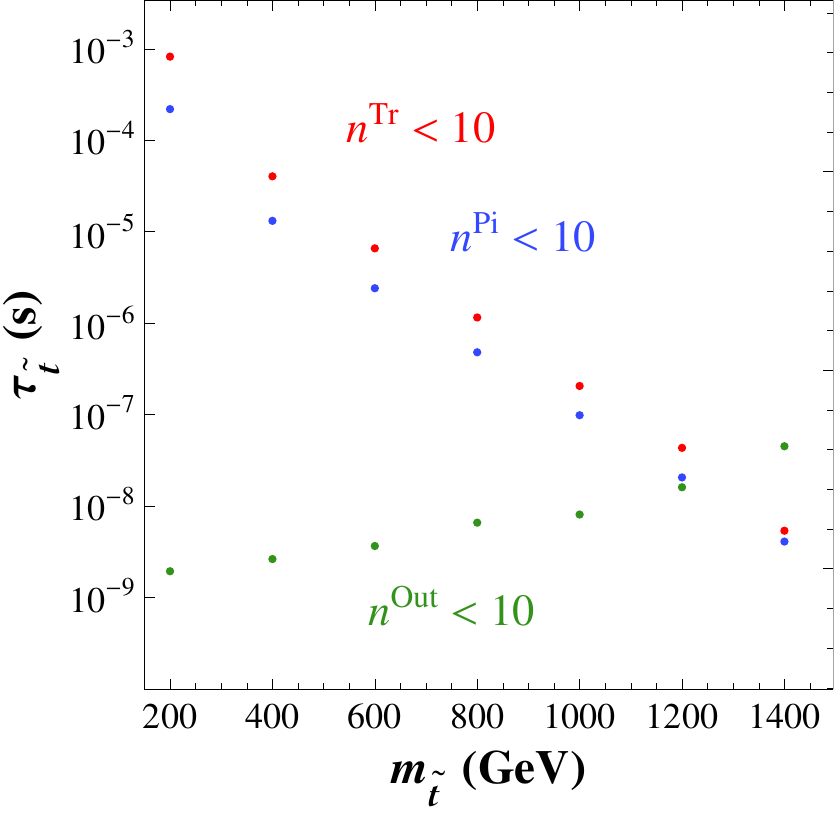}&
\includegraphics[scale=.93]{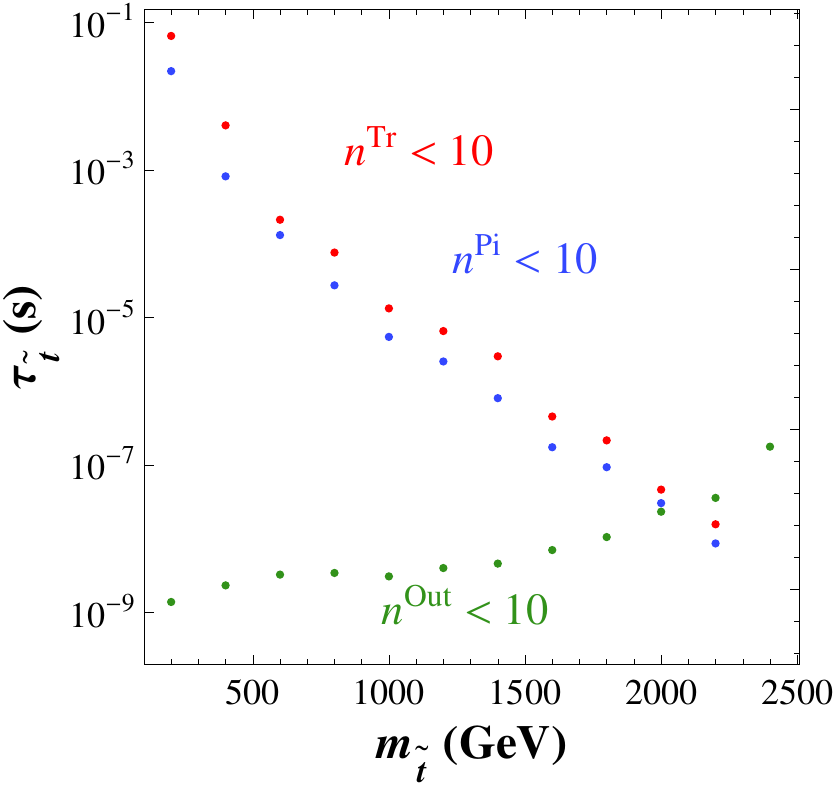}
\end{array}$
\end{center}
\vspace{-4mm}\caption{NLO LHC reach in the stop lifetime-stop mass 
plane at $L=25\,\mbox{GeV}$ in the left panel and $L=3000\,\mbox{GeV}$ 
in the right panel. In both panels, the Pixel reach is denoted by 
blue points and its interpolation is dashed while the Tracker reach 
is denoted by red points and its interpolation is dot-dashed. At last, 
the reach of metastable particles is denoted by green points and its 
interpolation is a solid line.}
\label{fig:MGL25&3000LTM}
\end{figure}

Now equipped with the stop decay length distributions, we can count
how many stops or anti-stops decay within the Pixel or Tracker in the
CMS detector at the integrated luminosity $L=25\; \mbox{fb}^{-1}$ and 
the maximum expected one $L=3000\;\mbox{fb}^{-1}$.
This allows us to obtain an estimate of the LHC reach in the
plane of stop mass versus lifetime with just two very simple
working hypotheses. 
First we neglect both the backgrounds of the SM and other SUSY 
particles than the lightest stop, which are respectively expected to have much 
shorter decay lengths and assumed to be too heavy to be produced at LHC.
Secondly we set the detector efficiencies to 100\% and declare $10$ decays inside 
one CMS detector part sufficient for the discovery of a displaced vertex
and $10$ decays outside for the discovery of a metastable stop. 
We require 10 decays instead of just 2 or 3 in order to reduce the numerical fluctuations 
and obtain a more stable numerical result. The analytical estimates in the next
section will allow to draw conclusions also for a different number of decays.
By doing so, the LHC reaches in Fig.~\ref{fig:MGL25&3000LTM} are obtained. 

Particularly, the Fig.~\ref{fig:MGL25&3000LTM} shows the MadGraph data-points 
in the stop mass-stop lifetime plane corresponding to $10$ decays inside Pixel 
(blue dots), Tracker (red dots) and outside the detector (green dots) at both integrated 
luminosities, with $L=25\;\mbox{fb}^{-1}$ shown in the left panel while 
$L=3000\;\mbox{fb}^{-1}$ in the right panel. Here, as in the next plots, 
the data-points for the stop decays outside the detector are always labeled by "Out".
Comparing the Pixel and Tracker reach, we are not surprised to see 
that they are pretty close to each other since their sizes are similar.
The tracker detector is larger than the pixel one and this gives a slightly
better reach, on the other hand we expect that the pixel detector would
offer a better precision and efficiency in the measurement of a displaced
vertex, so that this may overcome the geometrical advantage in a full
detector simulation. It is interesting that the stop kinematical distribution is such 
that a similar number of events is often obtained in the two detector parts, 
allowing on one side for a cross-check and on the other to better disentangle 
a long-lived stop from any SM background like B-mesons decaying mostly in 
the pixel detector.

We see clearly from the plot that the searches for displaced vertices and escaping
particles are complementary: the first covers the lower plane corresponding to 
short lifetimes, while the latter is mostly sensitive to the long lifetimes.
Combining the two searches it is possible to cover the parameter space
for any lifetime up to a maximal mass where the production cross-section starts 
to become too small to produce a sufficient number of stops.
Assuming that the LHC does not observe any signal, we could
therefore obtain a lifetime-independent lower limit on the stop mass 
at around $m_{\tilde t} \simeq 1300\,\mbox{GeV}$ for the 
integrated luminosity $L=25\,\mbox{fb}^{-1}$ and at 
around $m_{\tilde t} \simeq 2100\,\mbox{GeV}$ for the integrated 
luminosity $L=3000\,\mbox{fb}^{-1}$.

\subsection{Semi-analytic approximate analysis}

The semi-analytic analysis is realized via analytical estimates for the
decay length of the long-lived stop particles. Even in this 
case we use MadGraph 5 to compute the production cross section 
$\sigma$ and, therefore, the number of generated stop particles 
$N_0$ at LHC, from the product of the cross-section times the 
integrated luminosity ($\sigma L$). 
We complement the previous analysis with this semi-analytic approach 
in order to have a better control of the physical parameter space, faster
results and, at the same time, a useful check of the results of MadGraph 5.

The semi-analytic analysis is based on the well-known exponential decay 
formula for a particle travelling in a straight line, giving  the probability $P(d)$ 
that a particle decays at $d$. It reads 
\begin{equation}
 \label{Exponential_Decay}
 P(d)= \frac{\Gamma}{\beta\gamma c}\,\exp\left\{-\frac{\Gamma}
 {\beta\gamma c} d\right\}, 
\end{equation}
where $\Gamma$ is the decay rate in the centre of mass frame, $c$ 
the speed of light and $\beta\gamma$ is the relativistic $\beta\gamma$ factor, 
which is defined in term of the energy $E$ and the three-momentum 
$\vec{p}$ of the decaying particle as 
$\beta\gamma=(\left|\vec{p}\right\vert/E)/\sqrt{1-(\left |\vec{p}
\right \vert/E)^2}$. The factor in front of the exponential in Eq.~(\ref{Exponential_Decay}) is 
determined by the proper normalization of the probability $P(d)$, i.e.
the condition
\begin{eqnarray}
\label{Normaliz_Exp_dec_dim}
\nonumber\int\limits_{0}^{+\infty} P(d)\, dd = 1.
\end{eqnarray}
By using this exponential decay formula, the corresponding formula 
for the probability as function of a dimensionless coordinate $y$ 
can be very easily obtained. It reads 
\begin{equation}
 \label{decay_formula_dimensionless}
  P(y)=  \frac{1}{\beta\gamma}\,\exp\left\{-\frac{1}{\beta\gamma}\,y\right\}
\end{equation}
where $y = d\Gamma/c$ and  the normalization was obtained as for  $P(d)$. 
Here, we can explicitly see that any change in the decay rate $\Gamma$ 
can be always compensated by an appropriate change in the distance $d$ 
that the particle travels. Therefore, from the analytical expression 
of  $P(d)$ we can directly justify the rescaling procedure used for the
MadGraph events, that allowed us to cover the whole range of
decay rates from a single run.

Both the formula for $P(d)$ and $P(y)$ can be generalized to describe an 
exponential decay of a sample of particles, integrating over the particles
distribution in momentum and therefore $\beta\gamma $. 
To have a simpler expression, we will instead approximate such
integral by assuming a single effective value of $ \beta\gamma $
and by multiplying simply by the initial number of particles $N_0$. 
In this way, an estimate for the number of decaying particles of the sample 
as a function of $d$ and $y$, called $N(d)$ and $N(y)$ respectively, are 
achieved. They are 
\begin{eqnarray}
\label{Nr}
N(d)\!\!\!&=&\!\!\! N_0\, \frac{\Gamma}{\widetilde{\beta\gamma} c}\,\exp\left\{-
\frac{\Gamma}{\widetilde{\beta\gamma}c} d\right\},\\
\label{Ny}
N(y)\!\!\!&=&\!\!\!  N_0\, \frac{1}{\widetilde{\beta\gamma}}\,\exp\left\{-\frac{1}
{\widetilde{\beta\gamma}}\,y\right\},
\end{eqnarray}
where the coordinate $d = \sqrt{r^2 + z^2}$ can be taken as a function of the
coordinates in the CMS detector. We neglect here in first approximation
the bending of the trajectory by the magnetic field, which affects only
the case of stop hadronization into a charged hadron, or the energy loss
due to the interaction with the detector material, possibly negligible for 
a stop decay within the inner part of the detector.

\begin{figure}[t]
\begin{center}
$\begin{array}{cc}
\includegraphics[scale=0.96]{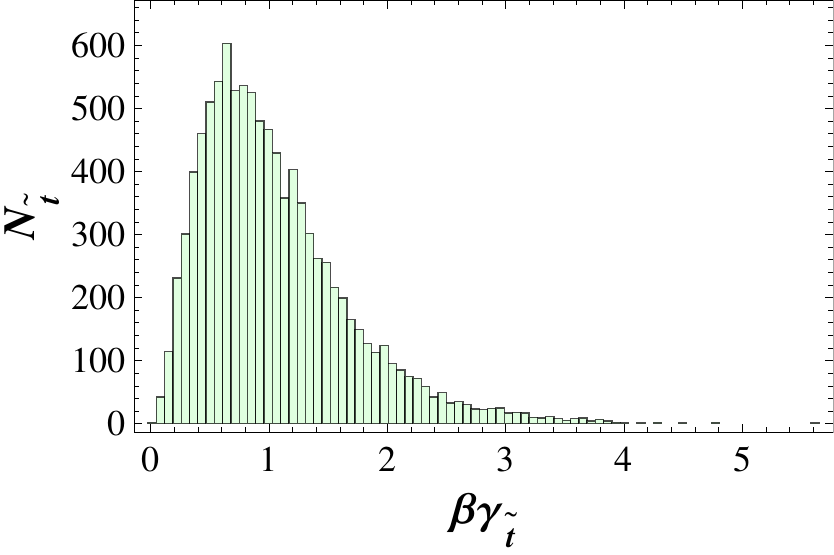}&
\end{array}$
\end{center}
\vspace{-4mm}\caption{MadGraph distribution of the factor $\beta\gamma $ 
for $m_{\tilde t}=800\,\mbox{GeV}$. }
\label{fig:BetaGammaDistr}
\end{figure}

To obtain the best approximation in Eqs.~(\ref{Nr}), (\ref{Ny}), we compute the 
stop $\beta\gamma$ distribution with MadGraph and compare the analytical 
formula above with different effective $\widetilde{\beta\gamma}$ with the decay
length's distribution obtained directly from the MadGraph run.
The $ \beta\gamma $ distribution is given in Fig.~\ref{fig:BetaGammaDistr} for a stop
mass of $800$ GeV. We see that even for relatively small mass, the stop
and anti-stop are mostly produced as non-relativistic, with a peak in the
$\beta\gamma $ distribution clearly below 1. 
We consider the analytical vertex distance distributions with different effective 
$\widetilde{\beta\gamma} $, i.e. taking the value at the maximum
$\beta\gamma^{max}\!=0.66$ or $(1/\beta\gamma)^{max}\!=0.8026$
or the average values $\langle\beta\gamma\rangle\!=0.9207$ and 
$\langle\langle1/\beta\gamma\rangle\rangle\!=\!1.24595$ .
For the reference mass $m_{\tilde t}=800\,\mbox{GeV}$ we compare
these distributions in the detector range $0\leqslant r(\mbox{m})\leqslant 17$ 
with the numerically computed distribution of displaced vertices 
and we select the value of $\widetilde{\beta\gamma} = \beta\gamma^{max}\!=0.66$ 
as the one giving the best fit. We compute then the arithmetic mean of the set 
of all the $\beta\gamma^{max}$-factors for all relevant masses and we
obtain the effective $\widetilde{\beta\gamma}$ as $\widetilde{\beta\gamma} = 0.66 $,
which we take from here on in the analytical equations  Eqs.~(\ref{Nr}),(\ref{Ny}). 
Such a value underestimates the distribution at large distances, so it gives a conservative 
estimate for the number of metastable particles decaying outside the detector.

\begin{figure}[t]
\begin{center}
$\begin{array}{cc}
\hspace{-3mm}\includegraphics[scale=.9]{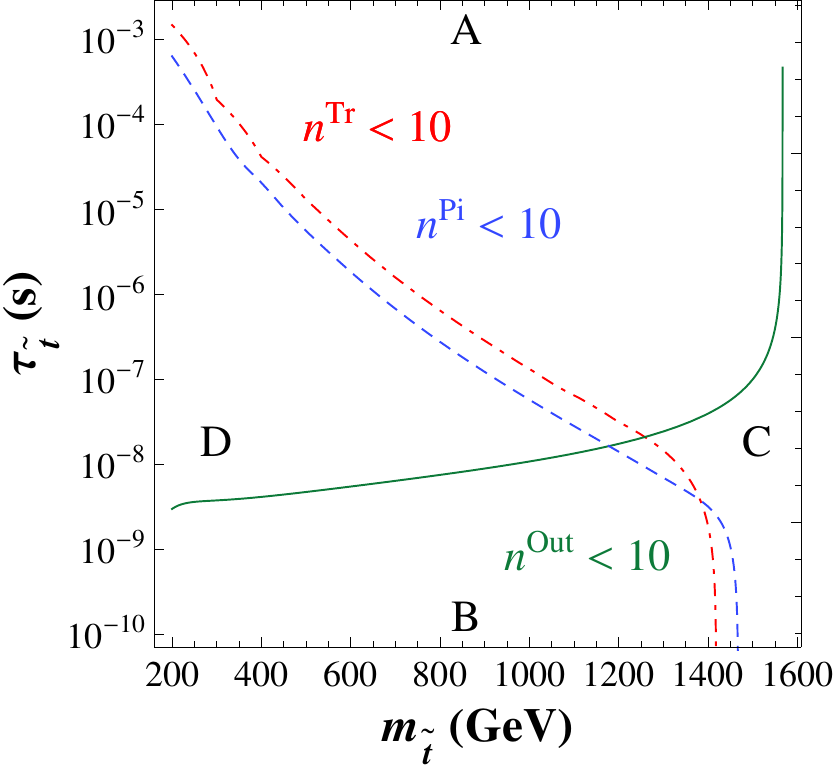}&
\includegraphics[scale=.9]{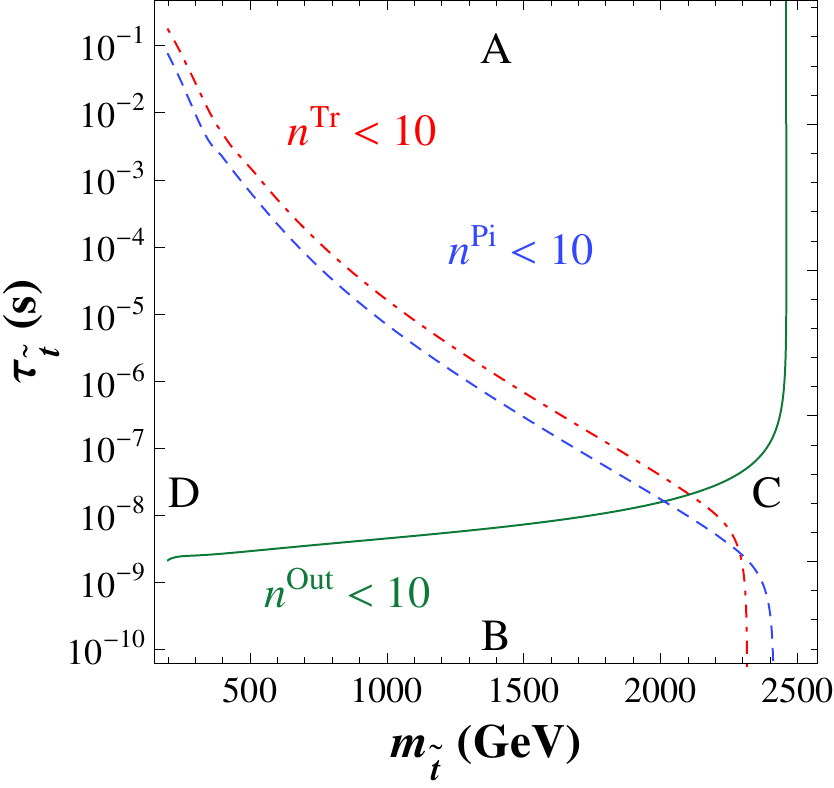}
\end{array}$
\end{center}
\vspace{-4mm}\caption{Semi-analytical LHC reach in the stop 
lifetime-stop mass plane at $L=25\,\mbox{fb}^{-1}$ in the left panel 
and $L=3000\,\mbox{fb}^{-1}$ in the right panel. In both panels, the 
Pixel reach is denoted by a blue dashed line while the Tracker reach 
by a red dot-dashed line. The reach of metastable particles is, instead, 
denoted by a green solid line.}
\label{fig:AnL25&3000LTM}
\end{figure} 
Let us work now out more details from the analytical formula of $N(d)$. 
Assuming just a spherical geometry for the detector parts, we can
integrate the expression with respect to $d$  from $r_i$ to $r_f$, which 
respectively stand for the initial and final radial distance from the IP of the part of interest 
of the detector, and from $r_f$ to $+\infty$ and so we have an analytical
approximate expressions for both number of stop decays that occur inside 
the detector parts and outside the detector.
They are given by the equations
\begin{eqnarray}
\label{formula_Inside_Detector}
N_{r_i\leqslant d\leqslant r_f}\!\!\!&=&\!\!\! N_0\left(\exp\left\{-\frac
{\Gamma}{\widetilde{\beta\gamma}}r_i\right\} - \exp\left\{-\frac{\Gamma}{\widetilde{\beta\gamma}}r_f\right\}\right),
\\
\label{formula_Outside_Detector}
N_{d\geqslant r_f}\!\!\!&=&\!\!\! N_0\exp\left\{-\frac{\Gamma}
{\widetilde{\beta\gamma}}r_f\right\}.
\end{eqnarray}
Since the number of particles generated by proton-proton collisions is given by the product of 
cross-section times luminosity, $ N_0 = \sigma L $, and  a power-law formula for $\sigma (m_{\tilde t})$ 
can be obtained fitting the MadGraph data,  we can solve Eq.~(\ref{formula_Inside_Detector})  and 
Eq.~(\ref{formula_Outside_Detector}) for the stop lifetime as a function of  the stop mass imposing 
$N_{r_i\leqslant d\leqslant r_f}=10$ or  $N_{d\geqslant r_f}=10$, for the values of
$ r_i, r_f$ corresponding to the pixel or tracker in the CMS detector.
In this way we can obtain a simple estimate of the 
LHC reach in the plane stop lifetime versus stop mass that we can
apply at different luminosities and also different values of diplaced vertices or 
metastable tracks $n$.
Note that the first equation is a  transcendental equation and can be 
only solved numerically, while the second can be simply solved analytically. 

\begin{figure}[t]
\begin{center}
$\begin{array}{cc}
\hspace{-3mm}\includegraphics[scale=.915]{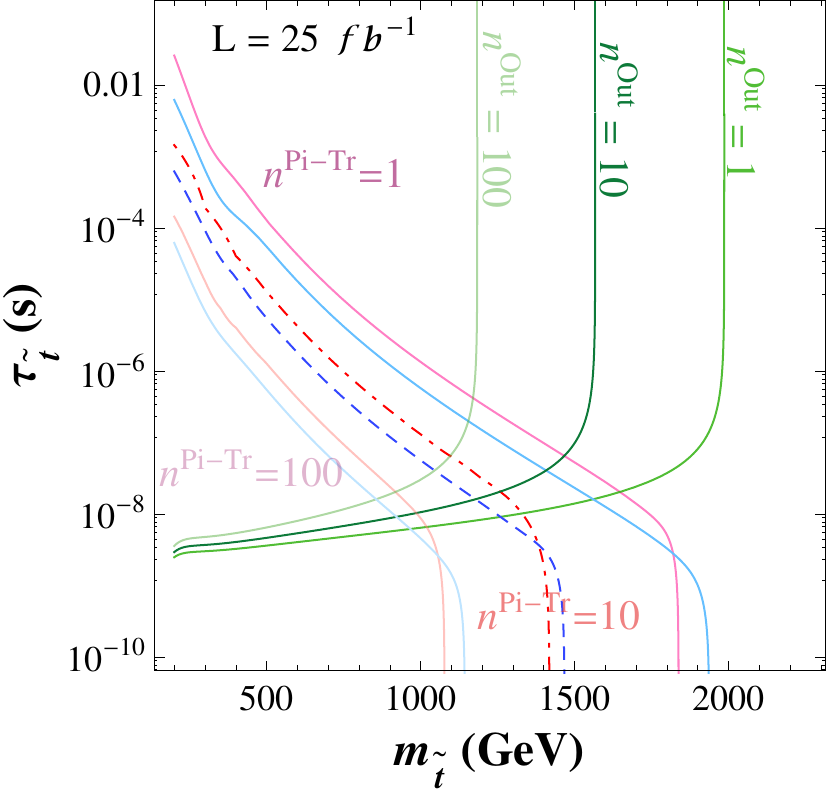}&
\includegraphics[scale=.90]{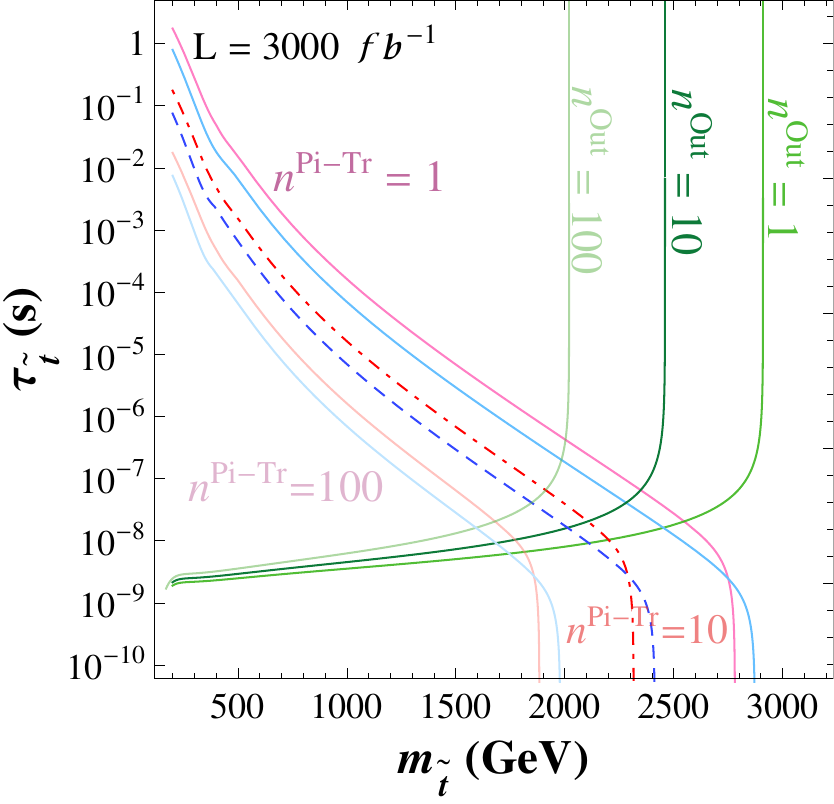}
\end{array}$
\end{center}
\caption{}
\label{fig:NumberEvents1&14}
\end{figure}

These results are plotted in Fig.~\ref{fig:AnL25&3000LTM}, where the 
dashed blue line, the dot-dashed red line and the green solid line denote, 
respectively, the Pixel, Tracker and Outside reach.   
Observing Fig.~\ref{fig:AnL25&3000LTM}, we can note that the crossing analytical curves 
identify in the plane four regions, which we have labelled $A,B,C$ and $D$. 
In the three regions A, B and D at least one signal is seen for any stop lifetime. 
In particular in the region D both types of signal are accessible, allowing 
to cross-check the measurement of the stop lifetime. Only in region $C$ 
no clear signal would be measured at the LHC. Of course such 
a region may be reduced by combining the two types of signals to have 
10 events in total or by loosening the requirement to fewer events.
To show the dependence on the requested number of vertices or metastable
tracks, we give in Fig.~\ref{fig:NumberEvents1&14} the LHC reach for  
$N=\left\{1, 10, 100\right\}$ again for both luminosities
$L=25\,\mbox{fb}^{-1}$ (left) and $L=3000\,\mbox{fb}^{-1}$ (right). 
It is clear here that the change in $N$ affects strongly the reach of
the metastable track search only at large masses, since for long lifetimes those 
events practically coincide to the total number of produced stops and anti-stops, 
which decreases very fast as a function of the stop mass.  For displaced
vertices, the change in $N$ just shifts the curves to a larger/smaller stop
lifetime.

\begin{figure}[!ht]
\begin{center}
$\begin{array}{cc}
\hspace{-3mm}\includegraphics[scale=.915]{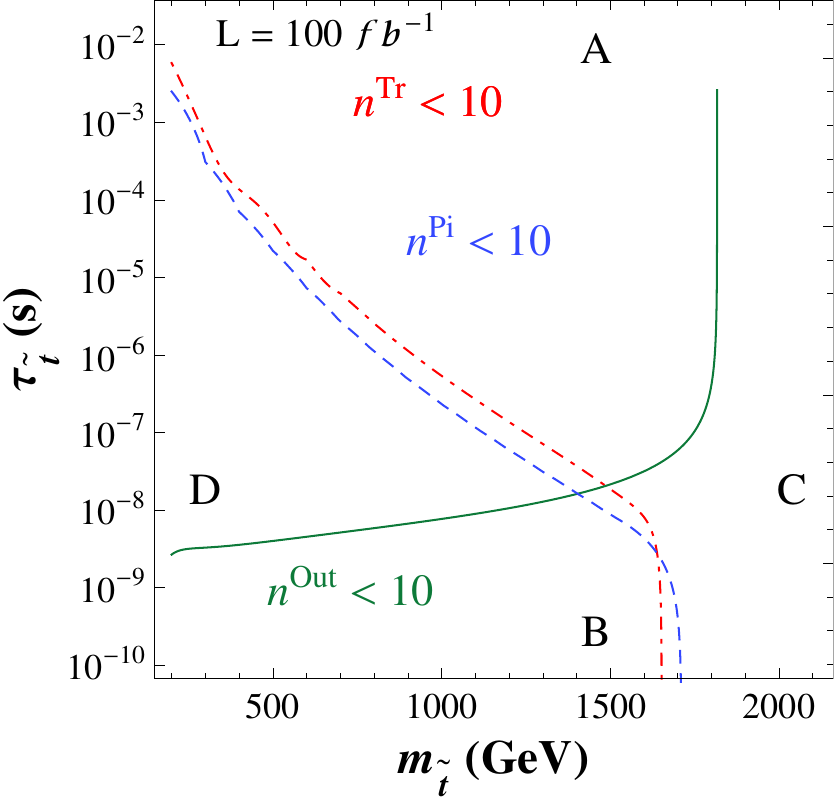}&
\includegraphics[scale=.90]{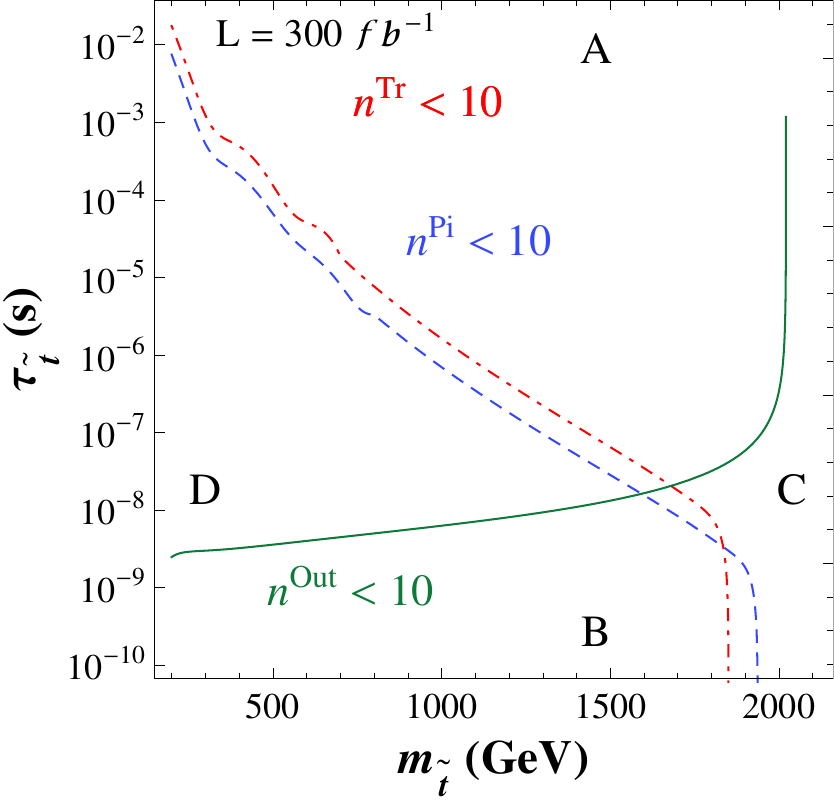}
\end{array}$
\end{center}
\caption{}
\label{fig:L100L300}
\end{figure}

Instead in Fig.~\ref{fig:L100L300} we give additional 
plots of the approximate reach for different LHC luminosities:
in the left panel we plot the LHC reach for  $L=100\,\mbox{fb}^{-1}$ whereas 
in the right one the LHC reach for $L=300\,\mbox{fb}^{-1}$. 
We see from the four different plots in Figs.~\ref{fig:AnL25&3000LTM} and \ref{fig:L100L300}
that the LHC has the chance to cover the whole parameter space in lifetime
up to stop masses of order $ 1300, 1500, 1700, 2100 $ GeV for a luminosity 
of $ 25, 100, 300, 3000 \,\mbox{fb}^{-1} $ respectively.

\newpage
\subsection{Comparison and discussion}

\begin{figure}[!ht]
\begin{center}
$\begin{array}{cc}
\hspace{-3mm}\includegraphics[scale=.913]{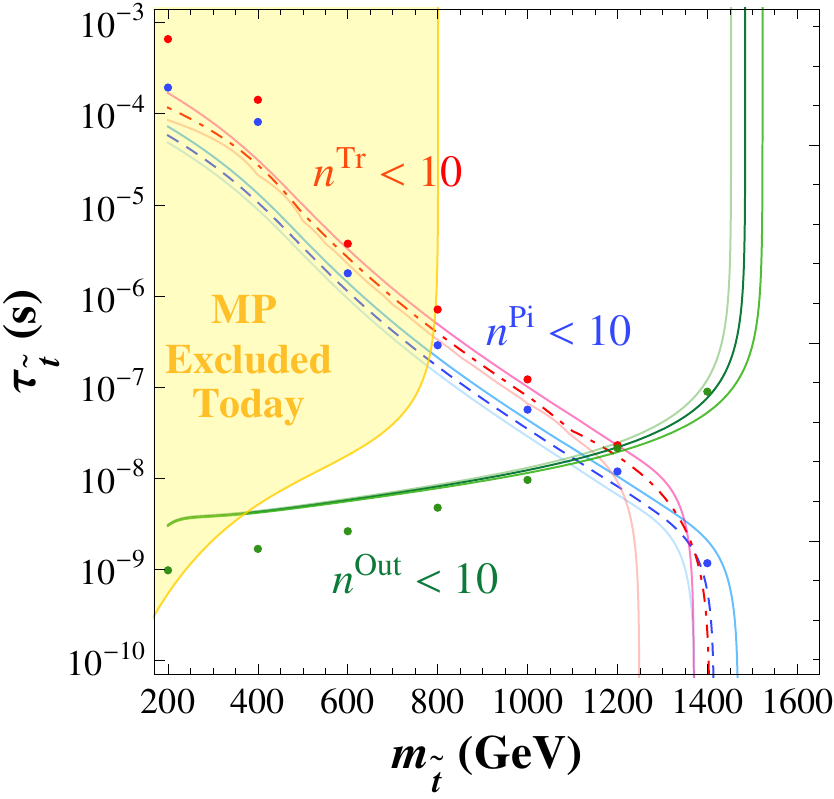}&
\includegraphics[scale=.90]{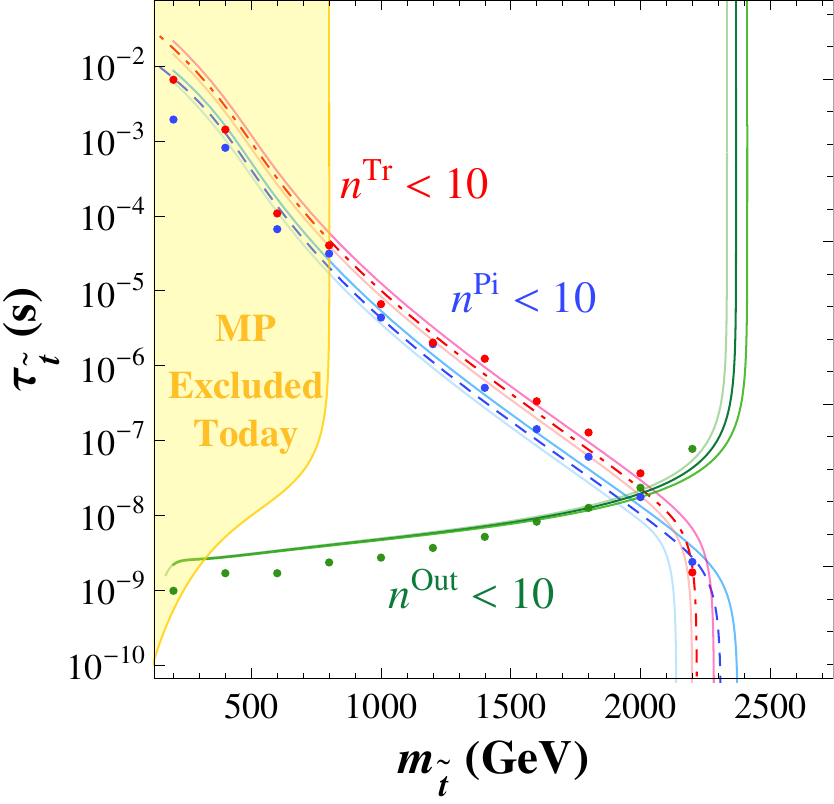}
\end{array}$
\end{center}
\caption{Comparison between the numerical LHC reach and the semi-analytical 
one at $L=25\;\mbox{fb}^{-1}$ (left panel) and $L=3000\;\mbox{fb}^{-1}$ 
(right panel). The centre of mass energy is $\sqrt{s}=14 \;\mbox{TeV}$ 
in both of them. In the upper left corner of each panel, the current 
excluded region for metastable particles (MP) is painted yellow. Each of 
the original analytical curves has their own uncertainty of $\pm 1\sigma$, 
represented by two new analytical curves which draw a region around the 
original curve for $n=10$.}
\label{fig:MGAnL25&L3000LTM}
\end{figure}

To compare directly the numerical and approximate LHC reach in the stop 
lifetime-stop mass plane, we plot all the curves together first without NLO 
correction in Fig.~\ref{fig:MGAnL25&L3000LTM}. 
Looking at this figure it is blatant to see a good agreement between 
the MadGraph data and the approximate curves at both integrated 
luminosities: $L=25\,\mbox{fb}^{-1}$, showed in the left panel, and 
$L=3000\,\mbox{fb}^{-1}$, showed in the right panel. The analytical
curves can be easily extended to consider $\pm 1\sigma$ statistical
error bars in the Poisson distribution, corresponding to $8$ and $12$
events respectively. We see that these curves give a very good description 
of the numerical data-points as they are mostly included in the $\pm 1\sigma$ band.
Only at low masses, where the statistics is large and the band becomes narrow,
 the analytical estimate deviates slightly from the numerical results.
 This is surely partially due to the spherical shape of the detector assumed in the 
 approximated expression, which causes a substantial error in the region with a 
 large number of events.
 In fact the data-points computed in the numerical analysis with the exact
 detector shape contain more signal events and give a wider reach in the 
 low mass region.
 
The current CMS excluded region for metastable particles (MP) 
\cite{CMS:collaboration_Long_Lived}, obtained at a centre of mass 
energy of $\sqrt{s}=8\,\mbox{TeV}$  and an integrated luminosity $L=18.8\,\mbox{fb}^{-1}$,
given by our analytical curve for zero decays outside the detector,
is also shown as a yellow region in the upper left corner of each panel.
We see that for long lifetimes the curve excludes stop masses below 800 GeV
and this coincide with the results obtained in \cite{CMS:collaboration_Long_Lived}.
Such bounds weakens for lifetimes below $ 10^{-7} $ s. Here it is worth highlighting 
that the current excluded region still leaves a considerable allowed region to be 
investigated at LHC in the future. 

\begin{figure}[!ht]
\begin{center}
$\begin{array}{cc}
\hspace{-3mm}\includegraphics[scale=.92]{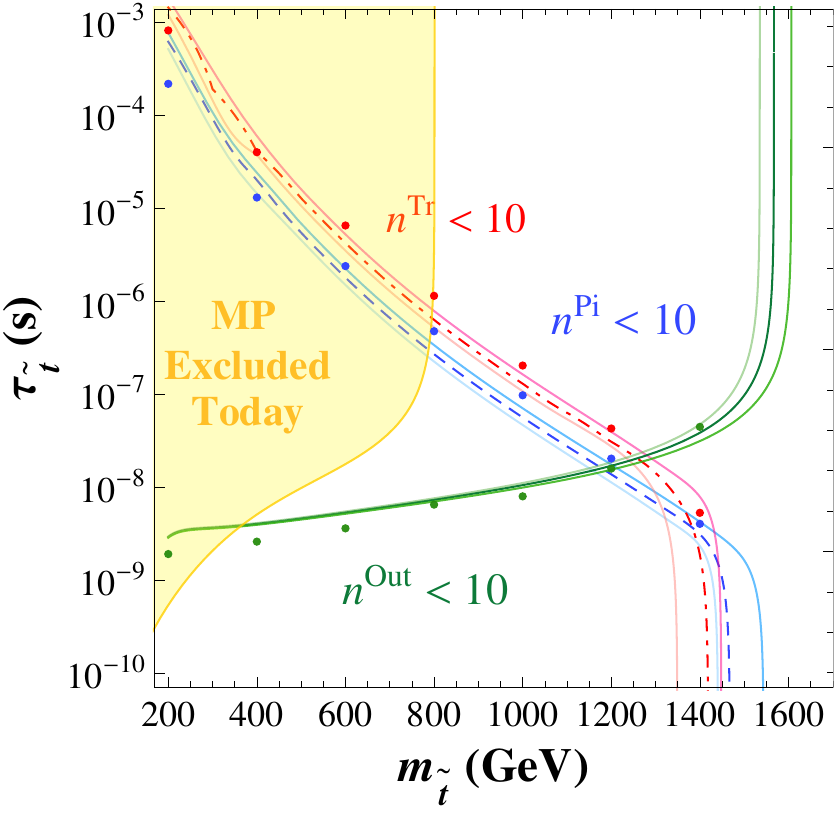}&
\includegraphics[scale=.92]{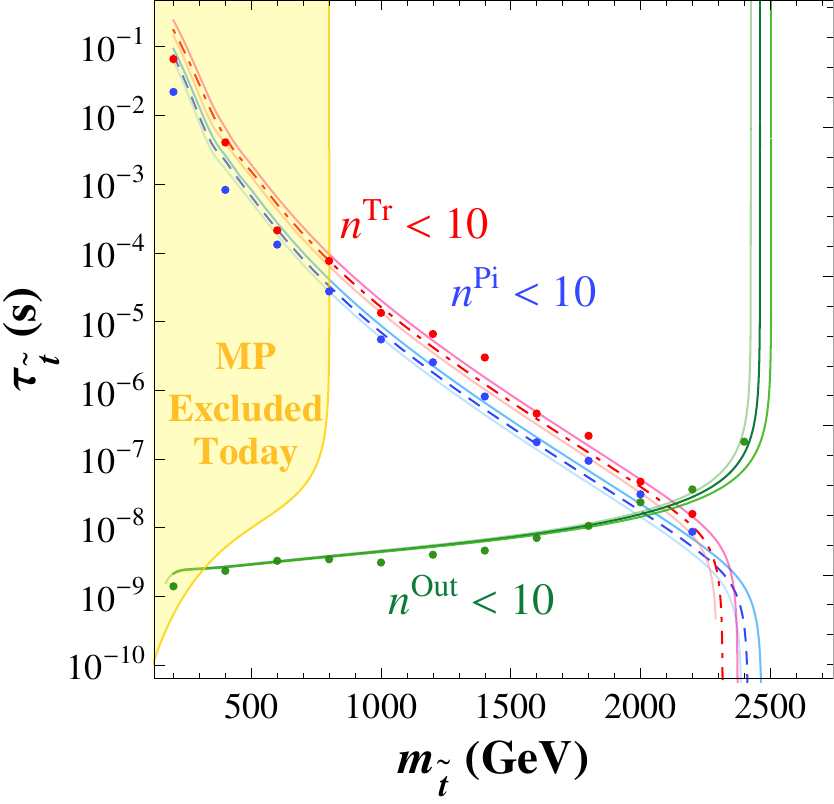}
\end{array}$
\end{center}
\vspace{-4mm}\caption{NLO LHC reach in the stop lifetime-stop mass 
plane at $L=25\,\mbox{GeV}$ in the left panel and $L=3000\,\mbox{GeV}$ 
in the right panel. In both panels, the Pixel reach is denoted by 
blue points and its interpolation is dashed while the Tracker reach 
is denoted by red points and its interpolation is dot-dashed. At last, 
the reach of metastable particles is denoted by green points and its 
interpolation is a solid line.}
\label{fig:MGL25&3000LTM-NLO}
\end{figure}

Finally we investigate as well the impact of NLO corrections to the 
stop production cross-sections. It is well known that processes involving
colored particles are strongly affected by QCD NLO corrections and
that those tend to increase the cross-section. It is therefore clear that
the LHC reach obtained with the LO cross-section is very conservative.
We redraw the same plots by multiplying the production cross-section
with the NLO k-factor of $1.6$ and we observe that such a correction
increases the yield substantially and extends the reach to larger stop masses. 
Indeed the NLO LHC reach in stop mass moves up by approximately
200 GeV, i.e. from 1200 to 1400 (2000 to 2200) GeV for  
25 (3000) $ fb^{-1}$ integrated luminosity.

\subsection{RPC and RPV models}
\label{Detector analysis of both RPC and RPV}

As we have already mentioned  beforehand, the discussion so far has been
centered on the presence of a displaced vertex or metastable tracks and is 
so independent from the particular stop decay channel, as long as 
it contains sufficiently many charged tracks to make the secondary vertex visible.
Therefore, our analysis is valid for both the RPC and the RPV stop decays 
we have discussed in Section~\ref{Stop coupling}. 
We can therefore interpret our results in terms of the two different 
model parameters.

\subsubsection{LHC reach for the RPC stop decay} 

\begin{figure}[t]
\begin{center}
$\begin{array}{cc}
\includegraphics[scale=0.944]{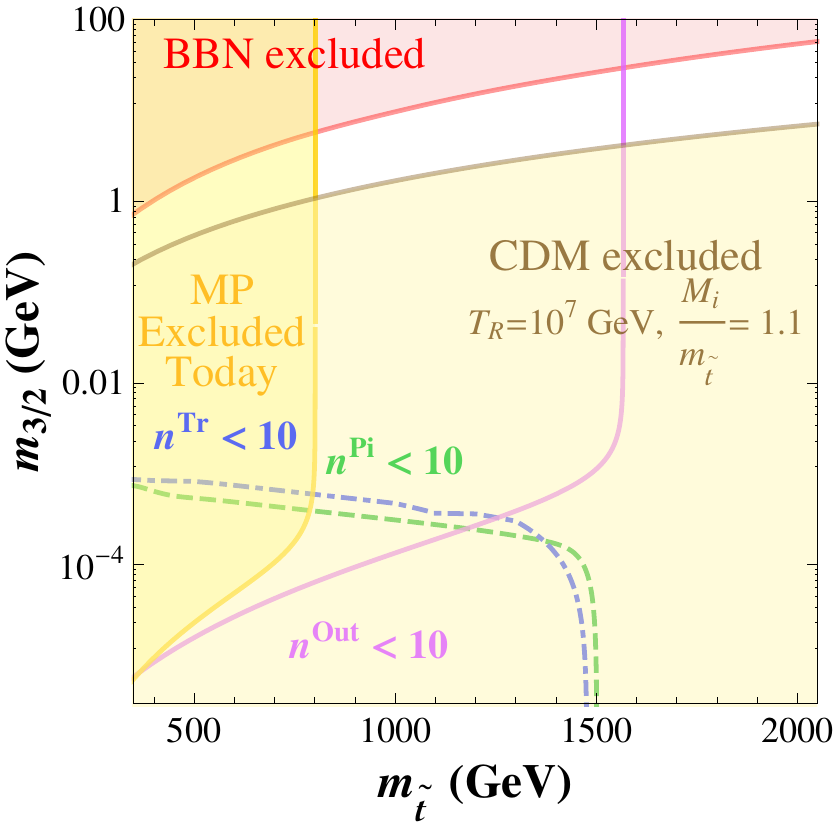}&
\includegraphics[scale=0.93]{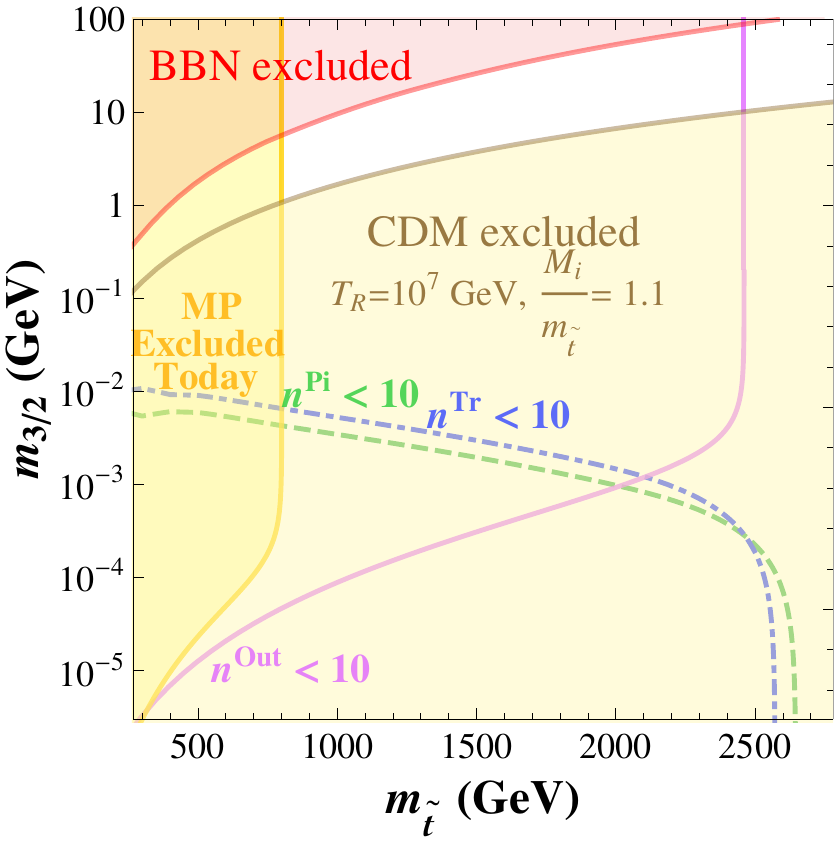}
\end{array}$
\end{center}
\caption{LHC reach in the stop mass-gravitino mass parameter space 
at NLO for the RPC decay $\tilde t \to \psi_{3/2} t$ for 25 $fb^{-1} $ (left panel) 
and for 3000 $fb^{-1} $ (right panel) luminosity. The 
centre of mass energy is $\sqrt{s}=14\,\mbox{TeV}$ in both of them. 
On the left side of each panel the current excluded region for 
metastable particles (MP) is tinted yellow. The analytical curves 
that correspond to 10 displaced vertices in Pixel, Tracker and 
outside the detector are denoted by a dashed green line, a 
dot-dashed blue line and a solid pink line rispectively.}
\label{fig:GravMass_vs_StopMass-NLO}
\end{figure} 

For the RPC stop decay into gravitino and top, the LHC reach can be easily reformulated 
in the plane $m_{\tilde t}\;\mbox{vs} \;m_{3/2}$ by using the analytical formula 
for the RPC stop lifetime, Eq.~(\ref{lifetime_RPC_stop_decay}).
In Fig.~\ref{fig:GravMass_vs_StopMass-NLO} we display these 
curves at NLO for Pixel, Tracker and the part outside the detector by means 
of a dashed green line, a dot-dashed blue line and a solid pink line 
respectively. Here we also plot the BBN excluded region (red region), 
the CDM excluded region (yellow region) and the 
current excluded region for metastable particles (yellow region). The 
CDM excluded region we have drawn is achieved for a reheating 
temperature of $T_R=10^7\,\mbox{GeV}$ and a ratio of physical gaugino 
masses to stop mass of $M_i/m_{\tilde t}=1.1$. 
The left panel of the Fig.~\ref{fig:GravMass_vs_StopMass-NLO} shows the LHC 
reach for the RPC stop decay for an integrated luminosity of 25 $fb^{-1} $ 
while the right panel for 3000 $fb^{-1} $.

We note that the cosmological allowed region 
(white region) appears above all of three detector curves, which 
means that stop NLSPs  with a consistent cosmology and high
reheat temperature can be detected only as metastable particles.  
The detection of a diplaced vertex would instead point to the
case of small gravitino mass and low reheating temperature.

\subsubsection{LHC reach for the RPV stop decay} 

For the bilinear RPV model, 
the LHC reach can be reformulated in the R-Parity breaking 
parameter-stop mass plane $\epsilon\;\mbox{vs}\;
m_{\tilde t}$ by using the analytic formula of the Bilinear RPV 
stop lifetime given by Eq.~(\ref{lifetime_RPV_Stop_decay}). 

\begin{figure}[t]
\begin{center}
$\begin{array}{cc}
\includegraphics[scale=0.898]{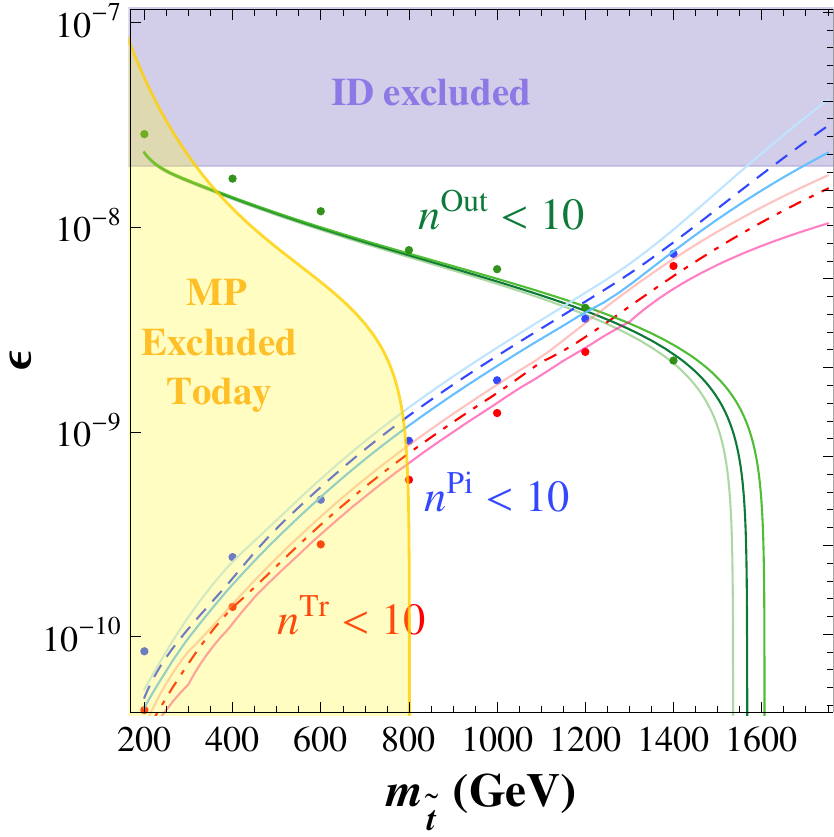}&\hspace{3mm}
\includegraphics[scale=0.898]{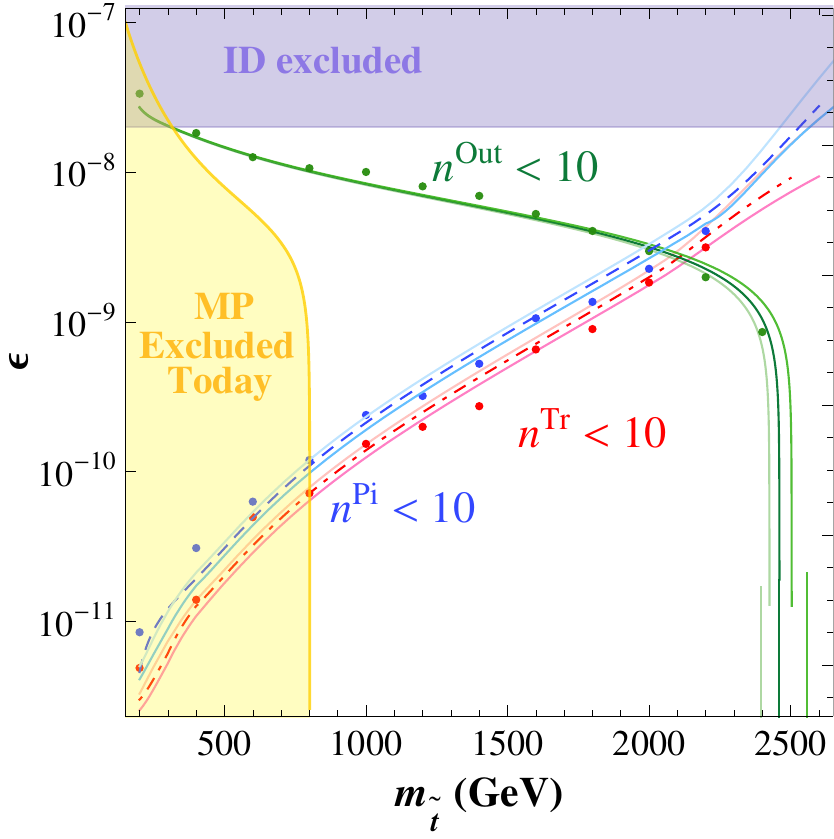}
\end{array}$
\end{center}
\vspace{-4mm}\caption{Reach in the R-Parity breaking parameter-stop mass plane 
for the RPV decay $\tilde t \to \ell^+ b$ for $25 fb^{-1} $ (left panel) and  for 
 $3000 fb^{-1} $ (right panel) including the NLO correction to the production
 cross-section. 
 At the lower left corner of each panel the current excluded region for 
metastable particles (MP) is tinted yellow. On the top of each panel, 
instead, the indirect indirect detection excluded region for the 
gravitino DM decay by Fermi-LAT collaboration is painted purpose.}
\label{fig:Epsilon_vs_Mass-NLO}
\end{figure}

In Fig.~\ref{fig:Epsilon_vs_Mass-NLO} the approximated curves at NLO for Pixel, 
Tracker and the part outside the detector are represented by a dashed 
blue line, a dot-dashed red line and a solid green line respectively. 
We take here a parameter point with $ \sin\theta/\cos\beta = 0.017$, 
corresponding to relatively small stop mixing. For larger stop
mixing the lifetime of the stop becomes shorter and therefore the
gap between the metastable particles bound at small $ \epsilon $
and the DM indirect detection bound increases. 

On the left panel the reach for 25 $fb^{-1}$ is shown whereas on the right 
panel for 3000 $fb^{-1}$ . Here we also display the corresponding 
MadGraph data by points which are blue for Pixel, red for Tracker and green 
for the decays that occur outside CMS.

At the lower left-hand corner of each plots the current excluded 
region for metastable particle (MP) is painted yellow. At the top 
of each panel, on the other hand, we see the indirect detection 
excluded region for the gravitino dark matter decay. The gravitino 
decay, indeed, leads to a diffuse $\gamma$-ray flux which can be 
compared to the $\gamma$-ray flux observed by Fermi-LAT collaboration 
\cite{FermiLat1:2010,FermiLat2:2010} so as to get a severe lower 
bound on the gravitino lifetime, and therefore a upper bound on 
the R-Parity breaking parameter $\epsilon$. Note that the latter
bound depends on the gravitino mass, but not the stop mass. 
Particularly, we take here the value of the upper bound on the R-Parity 
breaking parameter to be $\epsilon \simeq 2\times 10^{-8}$ for
a gravitino mass of the few GeVs from \cite{Vertongen:2011mu}. 

We see in this case that the analysis of displaced vertices is absolutely 
needed to close the gap between the indirect detection bound and
the possible metastable particle constraint. Indeed in this case both 
displaced vertices and metastable stops can be a signature in the
cosmologically favorable region.
Let us conclude this discussion pointing out that the majority of the 
parameter space is not excluded by the current indirect detection upper  
bound on the R-Parity breaking parameter for a decaying DM gravitino and 
neither by the current excluded region for metastable particles. 
The LHC experiment will be able in the near future to explore all the parameter
space up to the point where the stop NLSP is too heavy to be produced
in sufficient numbers.

\subsection{RPC and RPV stop decays at the LHC}
\label{Detector analysis of both RPC and RPV}

In the previous section we have performed a decay-channel independent analysis,
relying only on the presence or absence of a displaced vertex in the CMS detector.
In case one of such displaced vertices is observed, the question arises as to which
model is realized and if the decay is due to the RPC or RPV couplings.
In Fig.~\ref{2_body_decay} and Fig.~\ref{4_body_decay} we show the Feynman 
diagrams for these two decay channels of stop. 
\begin{figure}[t]
 \begin{minipage}[b]{8.5cm}
   \centering
   \begin{tikzpicture}[line width=1.3 pt,scale=1.6]
  \draw[fermion] (45:1)--(0,0);
  \draw[fermionbar] (315:1)--(0,0); 
  \draw[scalar] (180:1)--(0,0);
  \node at (-45:1.3) {$b$};
  \node at (45:1.3) {$\ell^+$};
  \node at (180:1.3) {$\tilde t$};	
\end{tikzpicture}
   \caption{2-body stop RPV decay.}
   \label{2_body_decay}
 \end{minipage}
 \begin{minipage}[b]{8.5cm}
  \centering
   \begin{tikzpicture}[line width=1.3 pt,scale=1.6]
  \draw[-][line width=2.6 pt] (30:1)--(0,0);
  \draw[fermionbar] (330:1)--(0,0);
  \draw[scalar] (180:1)--(0,0);
  \node at (-65:0.6) {$t$};
  \node at (30:1.4) {$\psi_{3/2}$};
  \node at (180:1.3) {$\tilde t$};
 \begin{scope}[shift={(0.85,-0.5)}]
	\draw[fermionbar] (30:1)--(0,0);
	\draw[vector] (-30:1)--(0,0);
	\node at (-75:0.7) {$W^+$};
	\node at (30:1.3) {$b$};	
\end{scope}
\begin{scope}[shift={(1.7,-1)}]
	\draw[fermion] (-30:1)--(0,0);
	\draw[fermionbar] (30:1)--(0,0);
	\node at (-30:1.3) {$\ell^+$};
	\node at (30:1.3) {$\nu_{\ell}$};	
\end{scope}
\end{tikzpicture}
   \caption{4-body stop RPV decay.}
   \label{4_body_decay}
 \end{minipage}
\end{figure}

From the point of view of the visible particles, both decays look similar,
since they correspond to a two-body RPV decay
\begin{equation}
 \label{RPVstopDecay}
 \tilde t\, \to \,b \,\ell^+ 
\end{equation}
and the four-body RPC stop decay:
\begin{equation}
\label{4_body_decay}
\tilde{t}\to\psi_{3/2} t \to \psi_{3/2} W^+ b \to \psi_{3/2} \,b\,\ell^+ 
\nu_{\ell}, 
\end{equation}
i.e. both decays end in a charged lepton and a b-quark as visible particles,
but they can be distinguished by the missing energy in the decay and
the particle kinematics.

The different phase space of these decay channels can be observed in 
different kinematical variables related to the visible particles. 
In particular, we focus our attention on the three following 
observables: the antilepton transverse momentum $P_{\ell T}$, the 
transverse mass of the pair antilepton-bottom $M_T$ and finally, the 
angle between the bottom and the antilepton momentum $\vartheta_{\ell b}$, 
which are defined by the formulas
\begin{eqnarray}
 \label{momentumTrans}
 P_{\ell T}\!\!\!&=&\!\!\!\sqrt{p_{\ell x}^2 + p_{\ell y}^2},
 \label{massTrans}
 \\M_T\!\!\!&=&\!\!\!\sqrt{(E_{\ell}+E_b)^2-(p_{\ell T}+p_{bT})^2},
 \label{angleDistr}
 \\ \vartheta_{\ell b} \!\!\!&=&\!\!\! \arccos\left(
 \frac{p_{\ell x} p_{b x} + p_{\ell y} p_{b y} + p_{\ell z} p_{b z}}
 {\sqrt{p_{\ell x}^2 + p_{\ell y}^2 + p_{\ell z}^2}
 \sqrt{p_{b x}^2 + p_{b y}^2 + p_{b z}^2}}\right).
\end{eqnarray}
The variables $E_{\ell}$, $\bold{p_{\ell}}=(p_{\ell x},p_{\ell y},
p_{\ell z})$, $E_{b}$, $\bold{p_{b}}=(p_{b x},p_{b y},p_{b z})$ stand 
for the energy and the three-momentum of antilepton and bottom, 
respectively. 

In order to compare these kinematical quantities for the two
decay channels of the stop,  we simulate both processes 
with MadGraph 5 for a stop mass of $m_{\tilde t}=800\,
\mbox{GeV}$ and the same stop decay rate 
of $\Gamma_{\tilde t}=2.02159\times 10^{-10}\,\mbox{GeV}$. 

\begin{figure}[t]
\begin{center}
$\begin{array}{cc}
\includegraphics[scale=0.85]{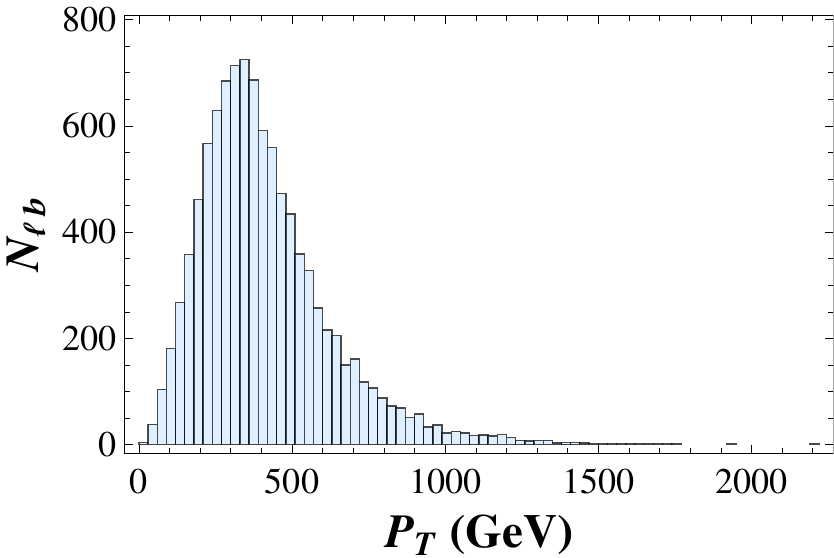}&
\hspace{3mm}\includegraphics[scale=0.9]{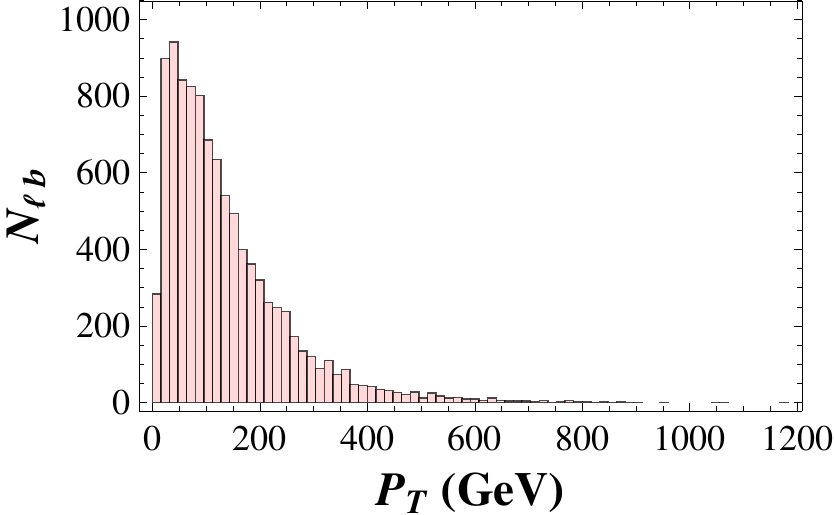}
\end{array}$
\caption{Transverse momentum distribution of the final antilepton for 
the two-body RPV $\tilde t$ decay (left) and four-body RPV $\tilde t$ 
decay (right) at $m_{\tilde t}=800\,\mbox{GeV}$.}
\label{fig:transv_momentum_distrs}
\end{center}
\end{figure}

\begin{figure}[t]
\begin{center}
$\begin{array}{cc}
\includegraphics[scale=0.84]{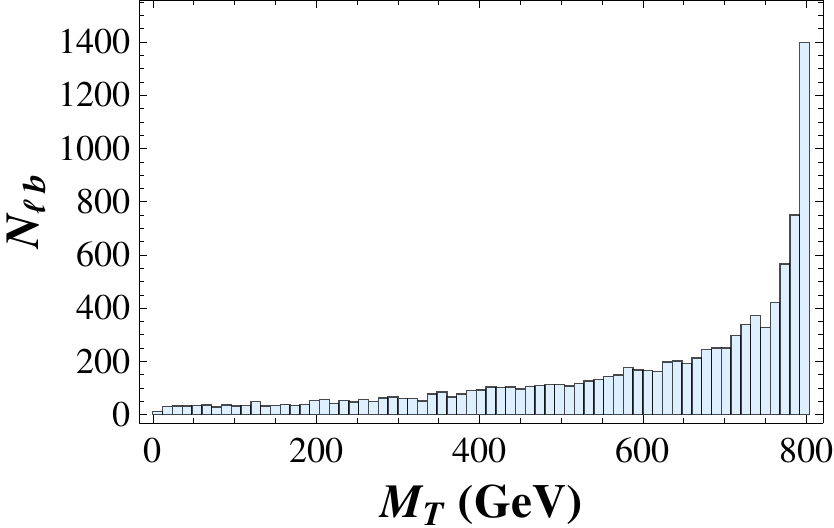}&
\hspace{5mm}\includegraphics[scale=0.84]{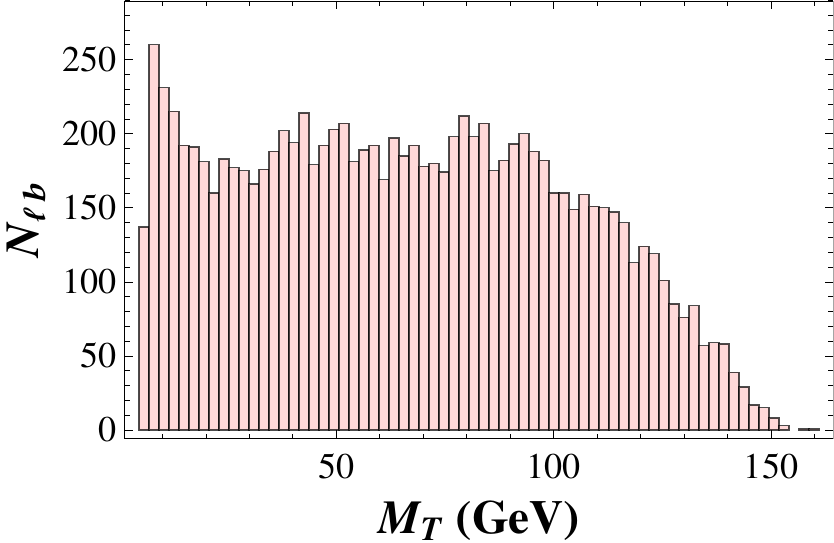}
\end{array}$
\caption{Transverse mass distributions of the final pair antilepton-bottom 
for the two-body RPV $\tilde t$ decay (left) and the four-body RPC $\tilde t$ 
decay (right) at $m_{\tilde t}=800\,\mbox{GeV}$.}
\label{fig:transv_mass_distrs}
\end{center}
\end{figure}

The transverse momentum distribution of the final antilepton for 
the two-body stop decay and the four-body stop decay are displayed in 
the left and the right panel of Fig.~\ref{fig:transv_momentum_distrs} 
respectively.  Looking at the figure we see that the peak of the two-body 
distribution is located at a much larger transverse momentum than the peak of 
the four-body distribution, because in the two body decay the lepton takes
away around half the rest energy of the stop.
Note that the presence of a lepton or an antilepton in the final 
state is very useful since it provides a clear and robust signature in the detector, 
which significantly suppresses the large Standard Model background (e.g. QCD). 
In addition, the measurement of light leptons at CMS is more precise than that 
of jets. 
The transverse mass distribution of the final pair antilepton-bottom 
for the two-body stop decay and the four-body stop decay are plotted 
in the left and the right panel of Fig.~\ref{fig:transv_mass_distrs} 
respectively. As we can see from the figure, the two distributions are 
very different. The two-body distribution has a peak and endpoint at a 
transverse mass of $M_T=800\,\mbox{GeV}$ corresponding to the stop
mass, while the four-body distribution shows no peak and tends to be 
concentrated to much smaller range. The difference is due to the missing 
energy in the four-body decay, while the two-body decay distribution allows
to infer the stop mass from the position of the end-point.

\begin{figure}[!ht]
\begin{center}
$\begin{array}{cc}
\includegraphics[scale=0.81]{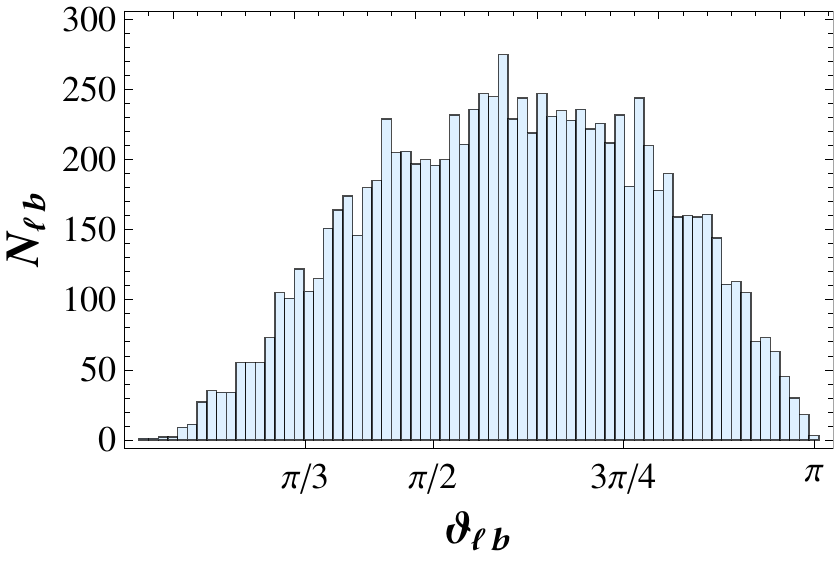}&
\hspace{7mm}\includegraphics[scale=0.80]{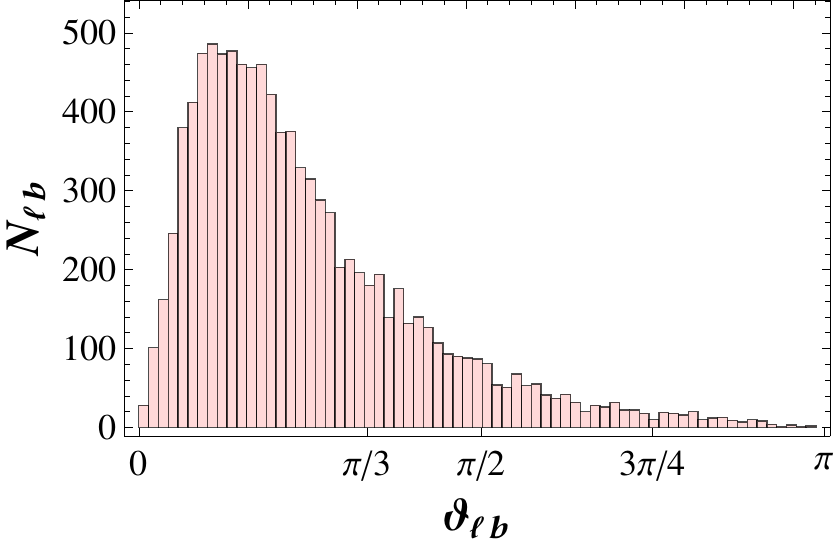}
\end{array}$
\end{center}
\caption{Angle distributions of the final pair antilepton-bottom for 
the two-body RPC $\tilde t$ decay (left) and four-body RPV $\tilde t$ 
decay (right) at $m_{\tilde t}=800\,\mbox{GeV}$.}
\label{fig:angl_distrs}
\end{figure}

The distribution of the angle between the final bottom momentum and the 
final antilepton momentum for the two-body stop decay and the four-body 
stop decay are displayed in the left and the right panel of 
Fig.~\ref{fig:angl_distrs} respectively. 
Observing the figure, we see that the two-body and the 
four-body distributions have no real peak. They are, instead, centered 
at different ranges of angle $\theta_{\ell b}$, which are $\pi/2
\lesssim \theta_{\ell b}\lesssim 3\pi/4$ for the two-body stop decay and 
$\theta_{\ell b}\lesssim \pi/3$ for the four-body distribution.  
This again is consistent with the fact that in the two-body decay the
lepton and bottom are back-to-back in the (slowly-moving) stop rest system, while
in the four-body decay they recoil against a non-negligible missing momentum.

In general these distributions can be used not only to distinguish between the two
RPC and RPV decays discussed here, but also to disentangle the signal from
the SM background from top decays and from the case  of a different model
or LSP, e.g. to compare to the case of a right-handed sneutrino LSP~\cite{deGouvea:2006wd}
or to other models like the $\mu\nu$MSSM~\cite{Choi-Fogliani:2010,Ghosh:2012pq}.

\subsection{Background and coincidence counting}

\begin{table}[t]
\centering
\begin{tabular}{|c||*{6}{c|}}
\hline
\multicolumn{7}{|c|}{$m_{\tilde t}=800\,\mbox{GeV}$ \;$\&$\;
$\tau_{\tilde t}= 3.254\times 10^{-9}\,\mbox{s}$ }\\
\hline
\hline
\backslashbox{\hspace{1.5mm}$\tilde t$}{\vspace{-2mm}$\tilde t^*$}
& \makebox[2em]{Bp} &\makebox[2em]{Pi}&\makebox[2em]{Tr}&\makebox[2em]{Ib}
&\makebox[2em]{Out}&\makebox[2em]{Tot}\\\hline\hline
\makebox[1.5em]{Bp} 
& $3.29\%$ & $5.90\%$ & $3.64\%$ & $1.20\%$ & $0.02\%$ & $14.05\%$ 
\\[.5ex]\hline\makebox[1.5em]{Pi} 
& $5.74\%$ & $17.13\%$ & $11.75\%$ & $5.08\%$ & $0.00\%$ & $39.70\%$ 
\\[.5ex]\hline\makebox[1.5em]{Tr} 
& $3.15\%$ & $11.98\%$ & $10.01\%$ & $5.05\%$ & $0.05\%$ & $30.24\%$ 
\\[.5ex]\hline\makebox[1.5em]{Ib} 
& $1.21\%$ & $5.09\%$ & $5.61\%$ & $3.89\%$ & $0.02\%$ & $15.82\%$ 
\\[.5ex]\hline\makebox[1.5em]{Out} 
& $0.03\%$ &$0.06\%$ & $0.05\%$ & $0.05\%$ & $0.00\%$ & $0.19\%$
\\[.5ex]\hline\makebox[1.5em]{Tot} 
& $13.42\%$ & $40.16\%$ & $31.06\%$ & $15.27\%$ & $0.09\%$ & $100\%$
\\[.5ex]\hline
\hline
\multicolumn{7}{|c|}{$m_{\tilde t}=800\,\mbox{GeV}$ \;$\&$\;
$\tau_{\tilde t}= 3.254\times 10^{-7}\,\mbox{s}$ }\\
\hline
\hline
\backslashbox{\hspace{1.5mm}$\tilde t$}{\vspace{-2mm}$\tilde t^*$}
& \makebox[2em]{Bp} &\makebox[2em]{Pi}&\makebox[2em]{Tr}&\makebox[2em]{Ib}
&\makebox[2em]{Out}&\makebox[2em]{Tot}\\\hline\hline
\makebox[1.5em]{Bp} 
& $0.00\%$ & $0.00\%$ & $0.00\%$ & $0.02\%$ & $0.18\%$ & $0.20\%$ 
\\[.5ex]\hline\makebox[1.5em]{Pi} 
& $0.00\%$ & $0.01\%$ & $0.01\%$ & $0.10\%$ & $0.77\%$ & $0.89\%$ 
\\[.5ex]\hline\makebox[1.5em]{Tr} 
& $0.00\%$ & $0.04\%$ & $0.05\%$ & $0.26\%$ & $1.41\%$ & $1.76\%$ 
\\[.5ex]\hline\makebox[1.5em]{Ib} 
& $0.01\%$ & $0.09\%$ & $0.17\%$ & $1.39\%$ & $8.95\%$ & $10.61\%$ 
\\[.5ex]\hline\makebox[1.5em]{Out} 
& $0.14\%$ &$0.73\%$ & $1.56\%$ & $8.63\%$ & $75.48\%$ & $86.54\%$ 
\\[.5ex]\hline\makebox[1.5em]{Tot} 
& $0.15\%$ & $0.87\%$ & $1.79\%$ & $10.40\%$ & $86.79\%$& $100\%$\\[.5ex]\hline
\end{tabular}
\caption{Coincidence counting for stop and antistop for $m_{\tilde t}=800\,\mbox{GeV}$ 
and lifetime $\tau_{\tilde t}= 3.254\times 10^{-9}, 10^{-7}\,\mbox{s}$. The labels Bp, Pi, Tr, 
Ib, Out stand respectively for the part of CMS before Pixel, Pixel, Tracker, the part 
between Tracker and the end of CMS and the part outside CMS. }
\label{mt800lifetime}
\end{table}
\vspace{6mm}

So far we have completely neglected the background of 
Standard Model and Supersymmetric particles, a clearly optimistic working hypothesis. 
The most important SM background  comes from the top-pair production at LHC. 
In fact, if the top quark decays into $W^{+}$ 
and bottom $b$ and at last, the weak boson $W^{+}$ decays into 
antilepton $\ell^{+}$ and neutrino $\nu_{\ell}$, we obtain in the final 
state the same visible particles as in the RPC and RPV stop decay. 
On the other hand, the top decays even before hadronizing and
much faster that the stop discussed here and therefore one can 
avoid all this SM background just by requiring a displaced vertex.
More difficult is to eliminate another source of background coming
from $ b \bar b Z \rightarrow b \bar b \ell^+ \ell^- $, where the b-decay
happens naturally away from the primary vertex and the lepton tracks 
are mis-reconstructed, as originating away from the interaction point.
Moreover also underlying events can give rise to particles pointing
to a secondary vertex, faking the presence of a long-lived particle.

In general a good strategy to eliminate such kind of reducible background
is to consider the presence of two displaced vertices in the same event,
both consistent with the same decay time. Indeed we can see that
in many cases one expects to have both stop and anti-stop to decay in
the same part of the detector and give a clear signal for the production
of two long-lived particles. We give in the table below the percentage
of decays of the stop and anti-stop in the different detector parts.
\begin{table}
\centering
\begin{tabular}{|c||*{6}{c|}}
\hline
\multicolumn{7}{|c|}{$m_{\tilde t}=2000\,\mbox{GeV}$ \;$\&$\;
$\tau_{\tilde t}= 3.254\times 10^{-9}\,\mbox{s}$ }\\
\hline
\hline
\backslashbox{\hspace{1.5mm}$\tilde t$}{\vspace{-2mm}$\tilde t^*$}
& \makebox[2em]{Bp} &\makebox[2em]{Pi}&\makebox[2em]{Tr}&\makebox[2em]{Ib}
&\makebox[2em]{Out}&\makebox[2em]{Tot}\\\hline\hline
\makebox[1.5em]{Bp} 
& $2.69\%$ & $6.94\%$ & $3.35\%$ & $0.78\%$ & $0.01\%$ & $13.77\%$ 
\\[.5ex]\hline\makebox[1.5em]{Pi} 
& $6.47\%$ & $21.58\%$ & $13.44\%$ & $3.69\%$ & $0.00\%$ & $45.18\%$ 
\\[.5ex]\hline\makebox[1.5em]{Tr} 
& $3.00\%$ & $13.10\%$ & $10.22\%$ & $3.95\%$ & $0.00\%$ & $30.27\%$ 
\\[.5ex]\hline\makebox[1.5em]{Ib} 
& $0.60\%$ & $4.23\%$ & $3.72\%$ & $2.23\%$ & $0.00\%$ & $10.78\%$ 
\\[.5ex]\hline\makebox[1.5em]{Out} 
& $0.00\%$ &$0.00\%$ & $0.00\%$ & $0.00\%$ & $0.00\%$ & $0.00\%$
\\[.5ex]\hline\makebox[1.5em]{Tot} 
& $12.76\%$ & $45.85\%$ & $30.73\%$ & $10.65\%$ & $0.01\% $& $100\%$
\\[.5ex]\hline
\hline
\multicolumn{7}{|c|}{$m_{\tilde t}=2000\,\mbox{GeV}$ \;$\&$\;
$\tau_{\tilde t}= 3.254\times 10^{-7}\,\mbox{s}$ }\\
\hline
\hline
\backslashbox{\hspace{1.5mm}$\tilde t$}{\vspace{-2mm}$\tilde t^*$}
& \makebox[2em]{Bp} &\makebox[2em]{Pi}&\makebox[2em]{Tr}&\makebox[2em]{Ib}
&\makebox[2em]{Out}&\makebox[2em]{Tot}\\\hline\hline
\makebox[1.5em]{Bp} 
& $0.00\%$ & $0.00\%$ & $0.00\%$ & $0.04\%$ & $0.12\%$ & $0.16\%$ 
\\[.5ex]\hline\makebox[1.5em]{Pi} 
& $0.00\%$ & $0.04\%$ & $0.05\%$ & $0.12\%$ & $0.81\%$ & $1.02\%$ 
\\[.5ex]\hline\makebox[1.5em]{Tr} 
& $0.00\%$ & $0.02\%$ & $0.04\%$ & $0.20\%$ & $1.91\%$ & $2.17\%$ 
\\[.5ex]\hline\makebox[1.5em]{Ib} 
& $0.04\%$ & $0.09\%$ & $0.36\%$ & $1.86\%$ & $10.27\%$& $12.62\%$ 
\\[.5ex]\hline\makebox[1.5em]{Out} 
& $0.10\%$ & $0.67\%$ & $1.60\%$ & $10.20\%$ & $71.46\%$ & $84.03\%$
\\[.5ex]\hline\makebox[1.5em]{Tot} 
& $0.14\%$ & $0.82\%$ & $2.05\%$ & $12.42\%$ & $84.57\%$ & $100\%$\\[.5ex]\hline
\end{tabular}
\caption{Coincidence counting for stop and antistop. The labels Bp, Pi, Tr, 
Ib, Out stand respectively for the part of CMS before Pixel, Pixel, Tracker, the part 
between Tracker and the end of CMS and the part outside CMS. }
\label{mt2000lifetime}
\end{table}
\vspace{6mm}

We see from Tables \ref{mt800lifetime} and \ref{mt2000lifetime},
that even in the unfortunate case of lifetime around $ 10^{-9} $ s, which is
at the boundary of the metastable particle searches, we have quite a large
statistics of coincident events in the pixel and tracker, reaching approximately
50\% of the displaced vertex events, irrespectively of the stop mass.
For longer lifetimes, the coincidence of two metastable particles in the
same event takes over, reaching quickly a large statistics.
We can therefore conclude that requiring two coincident events does not
reduce the signal statistics substantially, while it would certainly suppress
the background from misidentification or underlying events.

\section{Conclusion}

We have studied in this paper the reach of the LHC in models with a stop
NLSP and gravitino LSP, both for the case of R-parity conservation and violation.
In both cases we expect the stop to have a long lifetime leading to the possibility
of displaced vertices or metastable particles at the LHC.

From the cosmological perspective, in case of high reheat temperature above
the electroweak scale, the RPC scenario seems to prefer gravitino masses
above 1 GeV and therefore lifetimes giving metastable stops, while for the
RPV case only indirect detection of gravitino DM decay gives an upper
bound on the RPV parameter $\epsilon $, still consistent with stop
decays within the detector.

We have performed a model-independent analysis of the two expected
signals: displaced vertices in pixel or tracker detectors and metastable tracks. 
Our analysis is based on the MadGraph event generator, which allows us
to simulate both the stop-anti-stop production and its subsequent
decays. We assumed here that most of the other colored supersymmetric particles
are much heavier than the stop and outside of the LHC reach and therefore
we considered only the stop direct production, compute by MadGraph at LO and
 corrected to include NLO with a constant k-factor.
For the decay length distribution in the detector we also devised a simple analytical 
estimate based on the decay formula and the stop momentum 
distribution, that matches very well the MadGraph data-points and allows for a much
easier exploration of the parameter space. 

We have seen that even for relatively long lifetimes a substantial number
of decays can happen in the tracker or pixel parts of the CMS detector.
They surprisingly present similar number of events and can both be
exploited to measure the short lifetime region, while the metastable
particle search extends the reach quite strongly at long lifetimes.
We have therefore shown that the two search strategies are complementary
and that both are needed in order to cover all the macroscopic stop lifetimes 
up to a certain stop mass. It is interesting to note that both searches,
either displaced vertices or metastable states run out of steam at a similar
value of the stop mass, where the production cross-section becomes too small.
In particular we obtain, neglecting the background, in the most conservative case 
a mass reach up to 1400 (2000) GeV at LO and 1600 (2200) GeV at NLO for LHC at 14 TeV 
and with an integrated luminosity of $25 fb^{-1} $ and $ 3000 fb^{-1} $ respectively. 
As expected this reach is much larger than the minimal one for a metastable stau
from Drell-Yan production 
as given e.g. in~\cite{Heisig:2011dr}, but smaller to the reach expected for a 
metastable gluino. Of course a full detector simulation is needed to confirm
our findings, but the prospects seem to be very favorable for a combined analysis 
of both displaced vertices and metastable particle searches as discussed here.

We have translated the LHC reach in the parameter space of two models with 
gravitino DM, either RPC or RPV. In the first model the region compatible with high 
reheating temperature leads only to the signal of HSCP and the observation of 
displaced vertices in the
first run II fase with 25 $fb^{-1}$ would directly point to $T_R \sim 10^4- 10^3 $ GeV.
In the second model instead the search for displaced vertices allows to
close the gap between metastable particle and indirect detection bounds 
at low stop masses.
In case displaced vertices are seen, the visible particle kinematics allows 
to distinguish between RPC and RPV decays from the presence or absence
of any missing energy. The distributions and the flavour of the final lepton,
together with the value of the stop lifetime will be crucial to disentangle the
particular model.

In general, due to the pair production of stop and anti-stop, we expect 
to see in a large fraction of the events the coincident presence of two
displaced vertices or two metastable particles in the same event, allowing
to disentangle the signal from reducible backgrounds connected to
b-decays or misidentification of overlapping events.

\section*{Acknowledgements}

We would like to thank G. Arcadi, G. Hiller and Y. Nir for useful discussions.

G.A. and L.C. acknowledge financial support by the German-Israeli Foundation for scientific 
research and development (GIF) and partial financial support  by the EU FP7 ITN Invisibles 
(Marie Curie Actions, PITN-GA-2011-289442). 



\begin{thebibliography}{10}

\bibitem{SUSY-DM}
For recent reviews see
  J.~Ellis and K.~A.~Olive,
  In *Bertone, G. (ed.): Particle dark matter* 142-163
  [arXiv:1001.3651 [astro-ph.CO]];
  R.~Catena and L.~Covi,
  arXiv:1310.4776 [hep-ph].
  
\bibitem{Freedman:1976}
D.~Z.~Freedman, P.~van Nieuwenhuizen and S.~Ferrara, 
Phys. Rev. D 13 3214 (1976). 

\bibitem{Deser:1976}
S. Deser and B. Zumino, 
Phys. Lett. B 62 335 (1976).

\bibitem{Gravitino-problem}
  S.~Weinberg,
  Phys.\ Rev.\ Lett.\  {\bf 48} (1982) 1303.
  M.~Y.~.Khlopov and A.~D.~Linde,
  Phys.\ Lett.\ B {\bf 138} (1984) 265.
  J.~R.~Ellis, D.~V.~Nanopoulos and S.~Sarkar,
  Nucl.\ Phys.\ B {\bf 259} (1985) 175.
  
\bibitem{Bolz:1998}
M.~Bolz, W.~Buchm\"{u}ller and M.~Pl\"{u}macher,  
Phys. Lett. B 443 209 (1998) [arXiv:hep-ph/9809381].

\bibitem{Kawasaki:(2008)}
M. Kawasaki, K. Kohri, T. Moroi and A. Yotsuyanagi, 
Phys. Rev. D 78 065011 (2008) [arXiv:0804.3745 [hep-ph]].


\bibitem{Barbier:2004ez}
For a review on R-parity violation see e.g.
  R.~Barbier, C.~Berat, M.~Besancon, M.~Chemtob, A.~Deandrea, E.~Dudas, P.~Fayet and S.~Lavignac {\it et al.},
  Phys.\ Rept.\  {\bf 420} (2005) 1
  [hep-ph/0406039].
  
  
\bibitem{Takayama:2008} 
F.~Takayama and M.~Yamaguchi,
Phys. Lett. B 485 388 (2000) 
[arXiv:hep-ph/0005214].

\bibitem{Buchmuller:2007ui}
  W.~Buchmuller, L.~Covi, K.~Hamaguchi, A.~Ibarra and T.~Yanagida,
  JHEP {\bf 0703} (2007) 037
  [hep-ph/0702184 [HEP-PH]].

\bibitem{FermiLat1:2010}
Fermi-LAT Collaboration, A. A. Abdo et al., 
Phys. Rev. Lett. 104, 091302 (2010), 1001.4836.

\bibitem{FermiLat2:2010}
Fermi-LAT Collaboration, A. A. Abdo et al., 
Phys. Rev. Lett. 104, 101101 (2010), 1002.3603.

  
\bibitem{Bobrovskyi:2010}
S.~Bobrovskyi, W.~Buchm\"{u}ller, J.~Hajer and J.~Schmidt, 
JHEP 10 061 (2010) [arXiv:1007.5007].

\bibitem{Vertongen:2011mu}
  G.~Vertongen and C.~Weniger,
  JCAP {\bf 1105} (2011) 027
  [arXiv:1101.2610 [hep-ph]].
  
\bibitem{Ishiwata:2008tp}
  K.~Ishiwata, T.~Ito and T.~Moroi,
  Phys.\ Lett.\ B {\bf 669} (2008) 28
  [arXiv:0807.0975 [hep-ph]].
  
\bibitem{Graham:2012}
P.~W.~Graham, D.~E.~Kaplan, S.~Rajendran and P.~Saraswat, 
JHEP 1207 (2012) 149, [arXiv:1204.6038v1 [hep-ph]].

\bibitem{Khachatryan:2011ts}
  V.~Khachatryan {\it et al.}  [CMS Collaboration],
  JHEP {\bf 1103} (2011) 024
  [arXiv:1101.1645 [hep-ex]].


\bibitem{Aad:2011yf}
  G.~Aad {\it et al.}  [ATLAS Collaboration],
  Phys.\ Lett.\ B {\bf 701} (2011) 1
  [arXiv:1103.1984 [hep-ex]].
  
\bibitem{Chatrchyan:2012sp}
  S.~Chatrchyan {\it et al.}  [CMS Collaboration],
  Phys.\ Lett.\ B {\bf 713} (2012) 408
  [arXiv:1205.0272 [hep-ex]].
  
\bibitem{Chatrchyan:2012dxa}
  S.~Chatrchyan {\it et al.}  [CMS Collaboration],
  JHEP {\bf 1208} (2012) 026
  [arXiv:1207.0106 [hep-ex]].
  
\bibitem{Aad:2013pqd}
  G.~Aad {\it et al.}  [ATLAS Collaboration],
  Phys.\ Lett.\ B {\bf 722} (2013) 305
  [arXiv:1301.5272 [hep-ex]]. 
  
\bibitem{CMS:collaboration_Long_Lived}
CMS Collaboration, 
JHEP 1307 (2013) 122, [arXiv:1305.0491v2 [hep-ex]]. 

\bibitem{Aad:2013gva}
  G.~Aad {\it et al.}  [ATLAS Collaboration],
  Phys.\ Rev.\ D {\bf 88} (2013) 112003
  [arXiv:1310.6584 [hep-ex]].

\bibitem{ATLAS_collaboration:2013}
ATLAS collaboration, 
"Search for long-lived, heavy particles in final states with a muon and a multi-track displaced vertex in 
proton-proton collisions at $\sqrt{s}= 8$ TeV with the ATLAS  detector", 
ATLAS-CONF-2013-092.

\bibitem{Meade:2010ji}
  P.~Meade, M.~Reece and D.~Shih,
  JHEP {\bf 1010} (2010) 067
  [arXiv:1006.4575 [hep-ph]].
  
\bibitem{Bobrovskyi:2011vx}
  S.~Bobrovskyi, W.~Buchmuller, J.~Hajer and J.~Schmidt,
  JHEP {\bf 1109} (2011) 119
  [arXiv:1107.0926 [hep-ph]].
  
\bibitem{Hirsch:2005ag}
  M.~Hirsch, W.~Porod and D.~Restrepo,
  JHEP {\bf 0503} (2005) 062
  [hep-ph/0503059].
  
\bibitem{Ghosh:2012pq}
  P.~Ghosh, D.~E.~Lopez-Fogliani, V.~A.~Mitsou, C.~Munoz and R.~Ruiz de Austri,
  Phys.\ Rev.\ D {\bf 88} (2013) 1,  015009
  [arXiv:1211.3177 [hep-ph]].
  
\bibitem{Covi:2010au}
  L.~Covi, M.~Olechowski, S.~Pokorski, K.~Turzynski and J.~D.~Wells,
  JHEP {\bf 1101} (2011) 033
  [arXiv:1009.3801 [hep-ph]].
      
\bibitem{Steffen:2008bt}
  F.~D.~Steffen,
  Phys.\ Lett.\ B {\bf 669} (2008) 74
  [arXiv:0806.3266 [hep-ph]].
  
\bibitem{Endo:2010ya}
  M.~Endo, K.~Hamaguchi and K.~Nakaji,
  JHEP {\bf 1011} (2010) 004
  [arXiv:1008.2307 [hep-ph]].
 
\bibitem{Heisig:2011dr}
  J.~Heisig and J.~Kersten,
  Phys.\ Rev.\ D {\bf 84} (2011) 115009
  [arXiv:1106.0764 [hep-ph]]. 
  
\bibitem{Lindert:2011td}
  J.~M.~Lindert, F.~D.~Steffen and M.~K.~Trenkel,
  JHEP {\bf 1108} (2011) 151
  [arXiv:1106.4005 [hep-ph]].
  
\bibitem{Heisig:2012zq}
  J.~Heisig and J.~Kersten,
  Phys.\ Rev.\ D {\bf 86} (2012) 055020
  [arXiv:1203.1581 [hep-ph]].
  
\bibitem{Heisig:2013rya}
  J.~Heisig, J\"or.~Kersten, B.~Panes and T.~Robens,
  arXiv:1310.2825 [hep-ph].
  
\bibitem{Covi:2007xj}
  L.~Covi and S.~Kraml,
  JHEP {\bf 0708} (2007) 015
  [hep-ph/0703130 [HEP-PH]].
  
\bibitem{Ellis:2008as}
  J.~R.~Ellis, K.~A.~Olive and Y.~Santoso,
  JHEP {\bf 0810} (2008) 005
  [arXiv:0807.3736 [hep-ph]].
  
\bibitem{Katz:2009qx}
  A.~Katz and B.~Tweedie,
  Phys.\ Rev.\ D {\bf 81} (2010) 035012
  [arXiv:0911.4132 [hep-ph]].
  
\bibitem{Figy:2010hu}
  T.~Figy, K.~Rolbiecki and Y.~Santoso,
  Phys.\ Rev.\ D {\bf 82} (2010) 075016
  [arXiv:1005.5136 [hep-ph]].
  
\bibitem{Roszkowski:2012nq}
  L.~Roszkowski, S.~Trojanowski, K.~Turzynski and K.~Jedamzik,
  JHEP {\bf 1303} (2013) 013
  [arXiv:1212.5587].

  
\bibitem{Bobrovskyi:2012dc}
  S.~Bobrovskyi, J.~Hajer and S.~Rydbeck,
  JHEP {\bf 1302} (2013) 133
  [arXiv:1211.5584 [hep-ph]].
  
\bibitem{DiazCruz:2007fc}
  J.~L.~Diaz-Cruz, J.~R.~Ellis, K.~A.~Olive and Y.~Santoso,
  JHEP {\bf 0705} (2007) 003
  [hep-ph/0701229 [HEP-PH]].

\bibitem{Berger:2008ti}
  C.~F.~Berger, L.~Covi, S.~Kraml and F.~Palorini,
  JCAP {\bf 0810} (2008) 005
  [arXiv:0807.0211 [hep-ph]].

\bibitem{Kohri:2008cf}
  K.~Kohri and Y.~Santoso,
  Phys.\ Rev.\ D {\bf 79} (2009) 043514
  [arXiv:0811.1119 [hep-ph]].

\bibitem{Alwall:2010jc}
  J.~Alwall, J.~L.~Feng, J.~Kumar and S.~Su,
  Phys.\ Rev.\ D {\bf 81} (2010) 114027
  [arXiv:1002.3366 [hep-ph]].

\bibitem{Kats-Shih:2011}
Y.~Kats, D.~Shih, 
 JHEP 08 (2011) 049, [arXiv:1106.0030v2 [hep-ph]].

\bibitem{Marshall:2014cwa}
  Z.~Marshall, B.~A.~Ovrut, A.~Purves and S.~Spinner,
  arXiv:1402.5434 [hep-ph].
  
\bibitem{Asano:2010ut}
  M.~Asano, H.~D.~Kim, R.~Kitano and Y.~Shimizu,
  JHEP {\bf 1012} (2010) 019
  [arXiv:1010.0692 [hep-ph]].

\bibitem{Brust:2011tb}
  C.~Brust, A.~Katz, S.~Lawrence and R.~Sundrum,
  JHEP {\bf 1203} (2012) 103
  [arXiv:1110.6670 [hep-ph]].

\bibitem{Papucci:2011wy}
  M.~Papucci, J.~T.~Ruderman and A.~Weiler,
  JHEP {\bf 1209} (2012) 035
  [arXiv:1110.6926 [hep-ph]].
  
\bibitem{Martin:2011}
S.~P.~Martin, "A Supersymmetry Primer", [arXiv:hep-ph/9709356v6].

\bibitem{Hall:2011aa}
  L.~J.~Hall, D.~Pinner and J.~T.~Ruderman,
  JHEP {\bf 1204} (2012) 131
  [arXiv:1112.2703 [hep-ph]].

\bibitem{Heinemeyer:2011aa}
  S.~Heinemeyer, O.~Stal and G.~Weiglein,
  Phys.\ Lett.\ B {\bf 710} (2012) 201
  [arXiv:1112.3026 [hep-ph]].
  
\bibitem{Arbey:2011ab}
  A.~Arbey, M.~Battaglia, A.~Djouadi, F.~Mahmoudi and J.~Quevillon,
  Phys.\ Lett.\ B {\bf 708} (2012) 162
  [arXiv:1112.3028 [hep-ph]].

\bibitem{Giudice:1998bp}
  G.~F.~Giudice and R.~Rattazzi,
  Phys.\ Rept.\  {\bf 322} (1999) 419
  [hep-ph/9801271].
  
\bibitem{Meade:2009}
 P. Meade, N. Seiberg, and D. Shih, 
Prog.Theor.Phys.Suppl.177 (2009) 143, [arXiv:0801.3278 [hep-ph]].

\bibitem{Buican:2009}
M.~Buican, P.~Meade, N.~Seiberg, and D.~Shih, 
JHEP 0903 (2009) 016, [arXiv:0812.3668 [hep-ph]].

\bibitem{Beenakker:1997ut}
  W.~Beenakker, M.~Kramer, T.~Plehn, M.~Spira and P.~M.~Zerwas,
  Nucl.\ Phys.\ B {\bf 515} (1998) 3
  [hep-ph/9710451].
  
  \bibitem{madgraph5:2011}
J.~Alwall, M.~Herquet, F.~Maltoni, O.~Mattelaer and T.~Stelzer, 
JHEP 1106(2011)128, arXiv:1106.0522v1 [hep-ph].

\bibitem{Ji-Mohapatra:2008}
X. Ji, R. N. Mohapatra, S. Nussinov and Y. Zhang, 
Phys. Rev. D 78 (2008) 075032 [arXiv:0808.1904 [hep-ph]].

\bibitem{Endo-Shindou:2009}
M. Endo and T. Shindou, 
JHEP 0909 (2009) 037 [arXiv:0903.1813 [hep-ph]].

\bibitem{Fileviez-Spinner:2009}
P. Fileviez Perez and S. Spinner, 
Phys. Rev. D 80 (2009) 015004 [arXiv:0904.2213 [hep-ph]].

\bibitem{Nikolidakis:2007fc}
  E.~Nikolidakis and C.~Smith,
  Phys.\ Rev.\ D {\bf 77} (2008) 015021
  [arXiv:0710.3129 [hep-ph]].

\bibitem{Csaki:2011ge}
  C.~Csaki, Y.~Grossman and B.~Heidenreich,
  Phys.\ Rev.\ D {\bf 85} (2012) 095009
  [arXiv:1111.1239 [hep-ph]].
  
  \bibitem{Gates:1999ei}
  S.~J.~Gates, Jr. and O.~Lebedev,
  Phys.\ Lett.\ B {\bf 477} (2000) 216
  [hep-ph/9912362].


\bibitem{Fairbairn:2006gg}
  M.~Fairbairn, A.~C.~Kraan, D.~A.~Milstead, T.~Sjostrand, P.~Z.~Skands and T.~Sloan,
  Phys.\ Rept.\  {\bf 438} (2007) 1
  [hep-ph/0611040].
  
\bibitem{Jedamzik:2009uy}
  K.~Jedamzik and M.~Pospelov,
  New J.\ Phys.\  {\bf 11} (2009) 105028
  [arXiv:0906.2087 [hep-ph]].
  
\bibitem{Kawasaki:2004qu}
  M.~Kawasaki, K.~Kohri and T.~Moroi,
  Phys.\ Rev.\ D {\bf 71} (2005) 083502
  [astro-ph/0408426].
  
\bibitem{Jedamzik:2006xz}
  K.~Jedamzik,
  Phys.\ Rev.\ D {\bf 74} (2006) 103509
  [hep-ph/0604251].
  
  \bibitem{Kusakabe:2009}
M.~Kusakabe, T.~Kajino, T.~Yoshida and G. J.~Mathews,
Phys.Rev.D80:103501, (2009) 
[arXiv:0906.3516v1 [hep-ph]].

\bibitem{Kohri:2001jx}
  K.~Kohri,
  Phys.\ Rev.\ D {\bf 64} (2001) 043515
  [astro-ph/0103411].
  
  \bibitem{Sommerfeld}
A.~Sommerfeld, {\it Atombau und Spektrallinien, Band 2}, Vieweg \& Sohn (1939); \\
A.~D.~Sakharov, Zh.\ Eksp.\ Teor.\ Fiz.\  {\bf 18}, 631 (1948) [Sov.\ Phys.\ 
Usp.\  {\bf 34}, 375 (1991)];
J.~S.~Schwinger, {\it Particles, sources, and fields. Vol. 2,} Addison-Wesley 
(1989) (Advanced book classics series).

\bibitem{Covi:1999}
L. Covi, J. E. Kim and L. Roszkowski, 
Phys.~Rev~ Lett. 82 (1999) 4180 [arXiv:hep-ph/9905212].

\bibitem{Feng:2003}
J. L. Feng, A. Rajaraman and F. Takayama, 
Phys.~Rev.~Lett.~91 (2003) 011302 
[arXiv:hep-ph/0302215].

\bibitem{Moroi-Murayama:1993}
T. Moroi, H. Murayama and M. Yamaguchi, 
Phys. Lett. B 303 (1993) 289.

\bibitem{Bolz:2007}
M.~Bolz, A.~Brandenburg and W.~Buchm\"{u}ller, Nucl.
Phys. B 606 518 (2001) [Erratum-ibid. B 790 (2008) 336] 
[arXiv:0012052v3 [hep-ph]].

\bibitem{Pradler:2006qh}
  J.~Pradler and F.~D.~Steffen,
  Phys.\ Rev.\ D {\bf 75} (2007) 023509
  [hep-ph/0608344].

\bibitem{Hall:2009bx}
  L.~J.~Hall, K.~Jedamzik, J.~March-Russell and S.~M.~West,
  JHEP {\bf 1003} (2010) 080
  [arXiv:0911.1120 [hep-ph]].
  
\bibitem{Cheung:2011nn}
  C.~Cheung, G.~Elor and L.~Hall,
  Phys.\ Rev.\ D {\bf 84} (2011) 115021
  [arXiv:1103.4394 [hep-ph]].
  
\bibitem{Ade:2013zuv}
  P.~A.~R.~Ade {\it et al.}  [Planck Collaboration],
  arXiv:1303.5076 [astro-ph.CO].

\bibitem{Komatsu:2011}
E.~Komatsu et al. [WMAP Collaboration], 
Astrophys. J. Suppl. 192 (2011) 18 [arXiv:1001.4538 [astro-ph.CO]].
   
\bibitem{Beenakker:1996ed}
  W.~Beenakker, R.~Hopker and M.~Spira,
  hep-ph/9611232.

\bibitem{CMS:2008}
 S.~Chatrchyan {\it et al.}  [CMS Collaboration],
  JINST {\bf 3} (2008) S08004.

\bibitem{feynrules:2013}
A.~Alloul, N.~D.~Christensen, C.~Degrande, C.~Duhr, B.~Fuks,
arXiv:1310.1921v1 [hep-ph], 
http://feynrules.irmp.ucl.ac.be/wiki/ModelDatabaseMainPage

 
\bibitem{deGouvea:2006wd}
  A.~de Gouvea, S.~Gopalakrishna and W.~Porod,
  JHEP {\bf 0611} (2006) 050
  [hep-ph/0606296].

\bibitem{Choi-Fogliani:2010}
K. Y. Choi, D. E. Lopez-Fogliani, C. Mu\~{o}z nand R. R. de Austri, 
JCAP 1003 (2010) 028 [arXiv:0906.3681 [hep-ph]].


    
   
  \end{thebibliography}
\end{document}